\documentclass[runningheads]{llncs}

\usepackage{graphicx}
\usepackage{makecell}
\usepackage{cite}
\usepackage{geometry}
\usepackage[table]{xcolor}
\usepackage{algorithm}
\usepackage{algpseudocode}
\usepackage{amsmath}
\usepackage{caption}
\usepackage{subcaption}
\usepackage{longtable}
\usepackage{amssymb}

\usepackage[figuresright]{rotating}
\usepackage{svg}

\begin{document}

\title {Explainable Machine Learning-Based Security and Privacy Protection Framework for Internet of Medical Things Systems}

\titlerunning{Explainable ML-Based Security for IoMT Systems}

\author{Ayoub Si-ahmed \inst{1} \and
Mohammed Ali Al-Garadi \inst{2} \and
Narhimene Boustia \inst{3}}
\authorrunning{S. Ayoub et al.}

\institute{Laboratoire LRDSI/SIIR, Blida 1 University, PROXYLAN SPA/Subsidiary of CERIST, Algeria
\email{si\_ahmed.ayoub@etu.univ-blida.dz} \and
Emory University, Atlanta, USA
\email{m.a.al-garadi@emory.edu} \and
Laboratoire LRDSI/SIIR, Blida 1 University, Algeria
\email{nboustia@gmail.com}}

\maketitle              

\begin{abstract}
The Internet of Medical Things (IoMT) transcends traditional medical boundaries, enabling a transition from reactive treatment to proactive prevention. This innovative method revolutionizes healthcare by facilitating early disease detection and tailored care, particularly in chronic disease management, where IoMT automates treatments based on real-time health data collection. Nonetheless, its benefits are countered by significant security challenges that endanger the lives of its users due to the sensitivity and value of the processed data, thereby attracting malicious interests. Moreover, the utilization of wireless communication for data transmission exposes medical data to interception and tampering by cybercriminals. Additionally, anomalies may arise due to human error, network interference, or hardware malfunctions. In this context, anomaly detection based on Machine Learning (ML) is an interesting solution, but it comes up against obstacles in terms of explainability and privacy protection.
To address these challenges, a new framework for Intrusion Detection Systems (IDS) is introduced, leveraging Artificial Neural Networks (ANN) for intrusion detection while utilizing Federated Learning (FL) for privacy preservation. Additionally, eXplainable Artificial Intelligence (XAI) methods are incorporated to enhance model explanation and interpretation. The efficacy of the proposed framework is evaluated and compared with centralized approaches using multiple datasets containing network and medical data, simulating various attack types impacting the confidentiality, integrity, and availability of medical and physiological data. The results offer compelling evidence that the FL method performs comparably to the centralized method, demonstrating high performance. Additionally, it affords the dual advantage of safeguarding privacy and providing model explanation while adhering to ethical principles.

\end{abstract}

\section{INTRODUCTION}
Internet of Things is a technology that revolutionizes the field of information science by incorporating sensors associated with objects to collect data. Its ability to obtain information anywhere and anytime has led to its integration into various sectors, including the healthcare domain known as IoMT. Equipped with sensors and actuators, medical devices facilitate continuous, remote, and real-time collection of physiological data, such as glucose levels, body temperature, and heart rate, allowing constant health monitoring. The enhancement in both the quantity and quality of the amassed data serves to optimize treatment efficacy, mitigate medical inaccuracies, and expedite early disease detection. This transformation means a transition from curative to preventive healthcare, which considerably increases the chances of patient recovery.

Moreover, leveraging the gathered health data, automatic treatments can be administered to patients with chronic conditions through actuators. For example, diabetic patients can receive automated insulin injections based on their blood glucose levels. Similarly, individuals with irregular heart rhythms can be administered electrical shocks through pacemakers, while those with neurological disorders can benefit from simulated brain activity via Deep Brain Implants, enhancing the overall well-being and quality of life for patients with chronic illnesses. The health information collected can be stored on cloud servers or hospital databases for in-depth analysis, harnessing the power of AI-assisted healthcare under the supervision of healthcare professionals, often referred to as Healthcare 4.0.

Notwithstanding the myriad benefits offered by IoMT, it faces significant security and privacy challenges, as evidenced by the alarming statistic that indicates a 77\% increase in malware attacks on IoT devices in the first six months of 2022 \cite{comptiaIotStats}. The utilization of wireless communication for transmitting data exposes it to Man-In-The-Middle (MITM) attacks. Furthermore, other sources of anomalies may affect data integrity due to human errors during data processing, network interference during transmission, or malfunctions occurring at the medical equipment level \cite{si2023survey}. These anomalies can compromise the integrity, confidentiality, and availability of crucial medical data. Such anomalies could lead to misdiagnoses and treatment errors, potentially resulting in tragic consequences.

To mitigate these concerns regarding anomalies, ML is proposed as a solution. Considering that IoMT systems generate large amounts of data, they can help ML models to distinguish between normal and abnormal behaviour. This capability facilitates the detection of anomalies and is effective against zero-day and new attacks. These detection systems can operate in real-time, and with advancements in Deep Learning (DL), the process of attribute selection and image processing \cite{zhang2023sunetFF} becomes automated. 

Despite their potential benefits, ML-based security solutions face several challenges that must be addressed to ensure their effective, ethical, and regulatory-compliant deployment. A significant limitation lies in the integration of IDS based on ML into centralized architectures. While such architectures streamline data processing and model training, they raise critical concerns about data privacy and security. Sharing sensitive information across a centralized system may violate user privacy, particularly in sectors like healthcare, where patient data is highly confidential. Moreover, the central node itself represents a single point of failure: if compromised, it could jeopardize the entire system, leading to catastrophic consequences. Additionally, centralized architectures are prone to latency issues, as all data must be transmitted to and processed by the central node. This can hinder network scalability, limit computational capacity, and create bottlenecks, especially as data volumes grow.

Another major challenge is the inherent opacity of many ML models, often referred to as 'black-box' models. These models lack transparency in their decision-making processes, making it difficult for stakeholders to understand how predictions or classifications are derived. This issue is particularly critical in regulated industries such as healthcare, where international standards like the Health Insurance Portability and Accountability Act (HIPAA) in the United States mandate explainability and transparency in outcomes. However, HIPAA is not the only standard to consider. International frameworks such as the General Data Protection Regulation (GDPR) in the European Union, ISO/IEC 27001 for information security management, and ISO/IEC 27701—an extension of ISO/IEC 27001 specifically focused on privacy information management—provide robust guidelines for data protection, privacy, and ethical AI deployment. The World Health Organization (WHO) has also emphasized the importance of transparency, explainability, and intelligibility in AI systems used in healthcare, highlighting that AI models must be interpretable by medical professionals, regulators, and patients to foster trust and ensure ethical deployment \cite{WHO2021}. ISO/IEC 27701 complements ISO/IEC 27001 by adding specific requirements for privacy management, making it an essential tool for organizations seeking to align their security practices with international best practices in data privacy. Adopting these standards is crucial for organizations operating globally, as they ensure compliance with diverse regulatory requirements and foster user trust.

Furthermore, the ethical implications of ML-based security solutions are often overlooked in current discussions. Considerations such as fairness, accountability, and bias mitigation are critical to ensuring that these technologies do not inadvertently harm individuals or communities. For example, biased training data can lead to discriminatory outcomes, while a lack of accountability mechanisms can make it difficult to assign responsibility for errors or misuse. Addressing these ethical challenges requires a multidisciplinary approach, involving not only technical solutions but also input from ethicists, policymakers, and end-users. By integrating ethical principles into the design and implementation of ML systems, organizations can develop more trustworthy and socially responsible solutions.

This article aims to introduce a framework to enhance the security of IoMT systems through the design of an IDS based on ML. The proposed solution involves the utilization of FL as a training methodology, allowing the sharing of locally trained model weights on end-devices instead of raw data. This  approach preserves data privacy in alignment with regulations such as GDPR and HIPAA, while the distributed nature of FL mitigates the single point of failure associated with a centralized structure. By opting to share model weights rather than raw data, numerous key challenges are addressed. This reduces bandwidth consumption and alleviates network congestion, thereby facilitating system scalability. Adopting FL enhances data confidentiality, reduces the potential risks associated with a centralized model, and increases overall system efficiency and robustness.

Furthermore, the integration of XAI enhances the transparency and interpretability of the ML models, enabling stakeholders—including patients, model designers, and regulators—to understand how decisions are made. By providing clear and understandable explanations of the model’s predictions, XAI ensures compliance with regulatory requirements that mandate explainability, such as those outlined in HIPAA, GDPR, WHO and ISO. Through transparent explanations, a proven track record of reliable performance, and the provision of detection history in percentage form, this framework cultivates trust among users.

From an ethical perspective, the proposed framework addresses critical concerns such as fairness, accountability, and bias mitigation. By leveraging FL, the framework
 ensures that sensitive data remains on local devices, reducing the risk of biased outcomes that can arise from centralized data collection. This decentralized approach allows for the inclusion of diverse datasets, promoting fairness and reducing the likelihood of discriminatory results. The transparency provided by XAI fosters accountability, as it becomes easier to identify and address potential biases or errors in the system. By offering intuitive and accessible insights, even non-technical users can access results and track the model’s performance over time. The ethical design of the framework ensures that it not only meets technical and regulatory standards but also aligns with societal values, promoting the responsible and equitable use of AI in healthcare and beyond.

The contributions of the proposed solution can be summarized as follows:

\begin{enumerate}
\itemsep=0pt

\item Proposal of an efficient IDS  based on DL for intrusion detection.

\item The proposed architecture is built upon a privacy-preserving FL framework that is designed to accommodate the operational characteristics of IoMT environments. By supporting dynamic client participation and distributed collaborative learning, the proposed architecture preserves sensitive medical data during training, avoids single points of failure, and enhances the scalability of healthcare IoMT infrastructures.

\item The integration of XAI methods enhances transparency and interpretability, ensures regulatory compliance, and aids model designers in optimizing performance, while the demonstrated effectiveness of the detection history strengthens user confidence in the system.

\item The evaluation of the solution proposed is conducted on four distinct datasets containing network and medical data, further validating its applicability and robustness.

\item The proposed solution's performance is thoroughly evaluated and compared with the centralized method, demonstrating its effectiveness.

\item The framework addresses ethical considerations such as fairness, accountability and bias mitigation through FL and XAI, ensuring privacy, transparency, and compliance with regulations.

\end{enumerate}

The article is structured to provide a logical progression from foundational concepts to empirical validation and contextual analysis. Section \ref{RELATED WORK} first establishes the essential background on IDS, ML, FL, and XAI, followed by a review of related work to identify specific research gaps in ML-based security for IoMT systems. Building on this foundation, Section \ref{METHODOLOGY FOR PROPOSED IDS MODEL} details the methodology for the proposed IDS model, beginning with a description of the IoMT system architecture and the local training process. It then elaborates on the integrated use of a FL process for secure, privacy-compliant aggregation and XAI techniques for interpretability, an architecture designed to inherently address the ethical imperatives of data protection, transparency, and accountability. Section \ref{EXPIREMENT SETUP AND RESULTS} validates the framework through a multi-faceted evaluation, first examining the impact of FL parameters, then applying SHAP for model interpretability, and finally comparing performance against a centralized baseline. To position the contribution within the field, Section \ref{Comparison with Previous Work} compares the framework’s effectiveness and advancements with prior studies. Broader implications, limitations, and insights derived from the results are synthesized in Section \ref{DISCUSSION}. Section \ref{CONCLUSION} concludes by summarizing key findings and outlining promising directions for future research.

\section{BACKGROUND AND RELATED WORK}\label{RELATED WORK}

\begin{sidewaysfigure}[!ht]
    \centering
    \includegraphics[
        width=0.98\textheight,
        height=0.98\textwidth,
        keepaspectratio
    ]{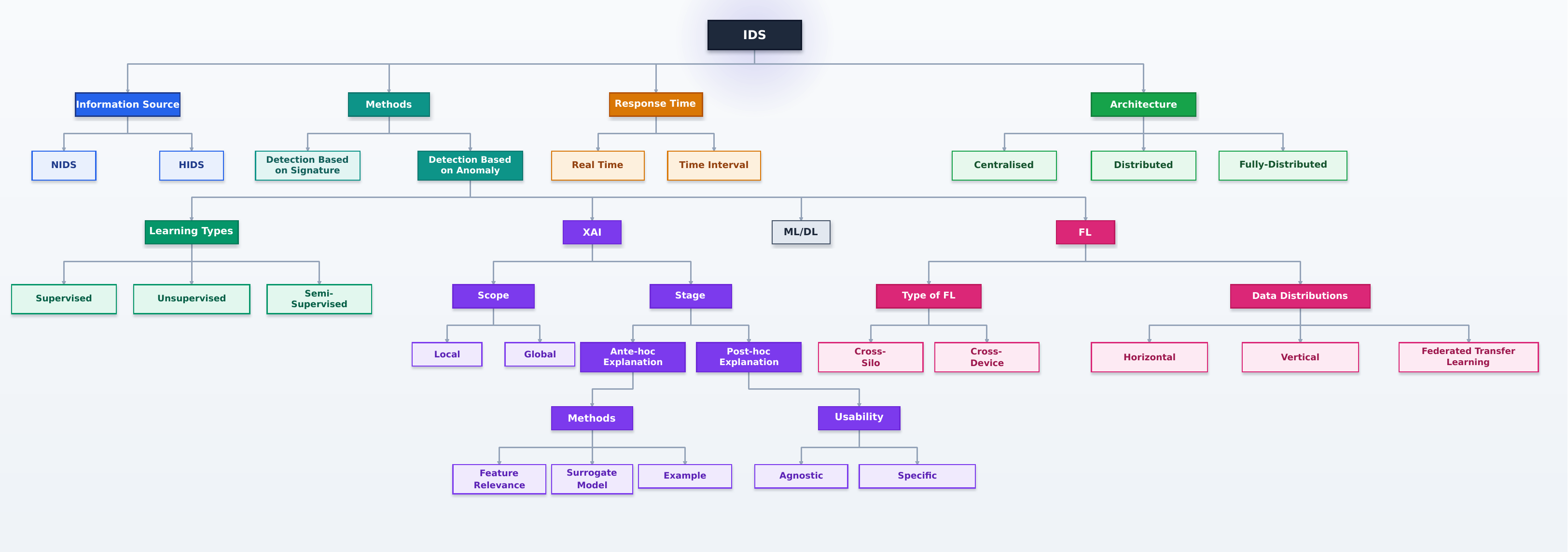}
    \caption{Flowchart Illustrating Fundamental Concepts Reviewed in Background.}
    \label{Background Schematic}
\end{sidewaysfigure}

In this section, the fundamental concepts of IDS, ML, FL, and XAI  are expounded upon, providing a comprehensive background. These concepts are visually represented in Figure \ref{Background Schematic}. Subsequently, the following section examines pertinent security solutions based on ML for anomaly detection within IoMT systems. The objective is to compare methods and identify the gaps upon which the proposed research is built. A detailed summary of all reviewed solutions is encapsulated in Table \ref{tbl1}.

\subsection{Background}

\newgeometry{margin=0.5cm}

\restoregeometry

The automation of surveillance and analysis within information systems critically relies on IDS. These systems are categorized by their data source: Network-based IDS (NIDS) monitor network traffic, while Host-based IDS (HIDS) analyze activities on individual hosts. Their analytical methods are broadly divided into two paradigms: Misuse Detection, which generates alerts by matching events against a database of predefined attack signatures, and Anomaly Detection, which leverages ML to model normal system behavior and flag significant deviations from it \cite{bace2001nist}. Architecturally, IDS can be implemented as Centralized, Fully-Distributed, or Partially-Distributed systems with hierarchical reporting mechanisms.

ML provides the foundational algorithms that empower modern anomaly-based detection. Standard ML paradigms include supervised learning, which trains models on labeled datasets to perform classification or regression tasks; unsupervised learning, which explores inherent patterns and structures within unlabeled data \cite{geron2017hands}; and semi-supervised learning, which leverages both labeled and unlabeled data to improve model performance and generalization where labeled data is scarce \cite{pise2008survey}. The application of these techniques is particularly salient for the IoMT, which generates vast volumes of complex and heterogeneous data. The capacity of ML to learn directly from data and adapt to evolving environments makes it highly suitable for detecting novel and sophisticated cyber threats in healthcare networks, a task where traditional static rule-based systems often prove inadequate.

FL has emerged as a distributed ML paradigm that directly addresses the core IoMT challenge of data privacy. In FL, a global model is iteratively trained by aggregating model updates from multiple participants without ever transferring or centralizing raw, sensitive patient data \cite{kairouz2021advances}. FL architectures are categorized based on the participant type into Cross-Silo FL, involving a few organizational entities, and Cross-Device FL, involving a large number of IoT devices. Furthermore, FL systems are classified based on data partitioning into Horizontal FL, where data across devices share the same feature space but concern different samples; Vertical FL, where data across entities contain different features but may overlap in sample identifiers; and Federated Transfer Learning, which extends transfer learning techniques to facilitate knowledge transfer across distributed data sources with limited common features or samples \cite{yang2019federated}. For IoMT, FL offers a significant advantage by enabling collaborative training on decentralized data silos, thus complying with stringent data protection regulations and facilitating the development of robust, generalized intrusion detection models that learn from a broader and more diverse data landscape than any single institution possesses.

\begin{longtable}{|c|l|}
    \hline
    \textbf{Notation} & \textbf{Description} \\
    \hline \hline
    $TP$ & True Positives \\
    \hline
    $FP$ & False Positives \\
    \hline
    $TN$ & True Negatives \\
    \hline
    $FN$ & False Negatives \\
    \hline
    $IoMT$ & Internet of Medical Things \\
    \hline
    $FL$ & Federated Learning \\
    \hline
    $XAI$ & eXplainable Artificial Intelligence \\
    \hline
    $IDS$ & Intrusion Detection System \\
    \hline
    $HIPAA$ & Health Insurance Portability and Accountability Act \\
    \hline
    $WHO$ & World Health Organization \\
    \hline
    $GDPR$ & General Data Protection Regulation \\
    \hline
    $ISO$ & International Organization for Standardization \\
    \hline
    $IEC$ & International Electrotechnical Commission \\
    \hline
    $ML$ & Machine Learning \\
    \hline
    $DL$ & Deep Learning \\
    \hline
    $ANN$ & Artificial Neural Network \\
    \hline
    $DNN$ & Deep Neural Network \\
    \hline
    $SVM$ & Support Vector Machine \\
    \hline
    $RF$ & Random Forest \\
    \hline
    $KNN$ & K-Nearest Neighbors \\
    \hline
    $CNN$ & Convolutional Neural Network \\
    \hline
    $LSTM$ & Long Short-Term Memory \\
    \hline
    $XSS$ & Cross-Site Scripting \\
    \hline
    $TTL$ & Time To Live \\
    \hline
    $U2R$ & User to Root \\
    \hline
    $R2L$ & Remote to Local \\
    \hline
    $ACCS$ & Australian Centre for Cyber Security \\
    \hline
    $EHMS$ & Enhanced Healthcare Monitoring System \\
    \hline
    $EMR$ & Electronic Medical Record \\
    \hline
    $EHR$ & Electronic Health Record \\
    \hline
    $CISO$ & Chief Information Security Officer \\
    \hline
    $SHAP$ & Shapley Additive Explanations \\
    \hline
    $MITM$ & Man-In-The-Middle \\
    \hline
    $CSV$ & Comma Separated Values \\
    \hline
    $AUC$ & Area Under the Curve \\
    \hline
    $ROC$ & Receiver Operating Characteristic \\
    \hline
    $LIME$ & Local Interpretable Model-Agnostic Explanations \\
    \hline
    $GRU$ &  Gated Recurrent Unit\\
    \hline
    $PCA$ & Principal Component Analysis \\
    \hline
    $AE$ & Auto Encoder\\
    \hline
    $LRGU$ & Logistic Redundancy Coefficient Upweighting \\
    \hline
    $MIFS$ & Mutual Information Feature Selection \\
    \hline
    $GWO$ & Grey Wolf Optimization \\
    \hline
    $SGD$ & Stochastic Gradient Descent \\
    \hline
    $MCPS$ & Medical Cyber-Physical Systems \\
    \hline
    $DoS$ & Denial of Service \\
    \hline
    $DDoS$ & Distributed Denial of Service \\
    \hline
    $HFL$ & Hierarchical FL \\
    \hline
    $SMOTE$ & Synthetic Minority Oversampling Technique \\
    \hline
    $DNS$ & Domain Name System \\
    \hline
    $SGRU$ & Sliced Gated Recurrent Unit \\
    \hline
    $sSAE$ & Stacked Sparse Autoencoder \\
    \hline
    $ICU$ & Intensive Care Unit \\
    \hline
    $TCP$ & Transmission Control Protocol \\
    \hline
    $UDP$ &User Datagram Protocol \\
    \hline
    $FTP$ &File Transfer Protocol \\
    \hline
    $SSH$ &Secure Shell \\
    \hline
    $HTTP$ &Hypertext Transfer Protocol  \\
    \hline
    $MLP$ & Multi-Layer Perceptron \\
    \hline
    $LR$ & Logistic Regression  \\
    \hline
    $RFE$ & Recursive Feature Elimination  \\
    \hline
    $NB$ & Naive Bayes  \\
    \hline
    $PSO$ & Particle Swarm Optimization  \\
    \hline
    $NIDS$ & Network-based IDS  \\
    \hline
    $HIDS$ & Host-based IDS  \\
    \hline
    $DT$ & Decision Tree  \\
    \hline
    $SRU$ & Simple Recurrent Unit \\
    \hline
    $Dintpkt$ & Destination Inter Packet \\
    \hline
    $dstjitter$ & Destination Jitter \\
    \hline
    $dstload$ & Destination Load \\
    \hline
    $srcload$ & Source Load \\
    \hline
    $dst\_port$ & destination ports\\
    \hline
    $src\_port$ & source port\\
    \hline
    \caption{Notation Table} 
\end{longtable}

Despite their high predictive efficacy, complex ML and FL models often operate as black boxes, creating a significant barrier to their adoption in high-stakes clinical settings.  XAI addresses this critical issue by providing methodologies to make model decisions understandable, interpretable, and trustworthy \cite{van2004explainable}. A comprehensive framework for  XAI can be described across three dimensions: explainability as the capacity to articulate the model's internal processes, interpretability as the ability to provide intuitive insight into the model's reasoning, and transparency as the inherent understandability of the model's mechanics \cite{islam2019infusing}. Models range from being intrinsically interpretable, such as Decision Trees (DT), to opaque, requiring post-hoc explanation techniques, such as Deep Neural Networks (DNN). This often creates a trade-off, where the superior performance of opaque models must be balanced against the need for explainability. Post-hoc methods operate on both local and global levels, employing techniques like feature relevance analysis, surrogate models, and representative examples to elucidate the model's behavior \cite{speith2022review}. In the context of IoMT security,  XAI is not a luxury but a necessity. It allows security analysts and healthcare professionals to audit model predictions, verify that alerts are based on clinically relevant features rather than data artifacts, and build the trust required for operational deployment. This capability is crucial for ensuring that automated decisions do not jeopardize patient safety and that the reasoning behind any alert can be rigorously validated, which is paramount for regulatory compliance and clinical adoption.

\subsection{Related Work}

The landscape of IDS for the IoMT is vast and can be systematically organized by its underlying architectural and methodological approaches. This section provides a critical review, first examining solutions that leverage ML and DL within a centralized framework, then exploring those that utilize FL to preserve data privacy, and finally, investigating emerging approaches that incorporate XAI for interpretability.

Most existing research uses centralized models where data are aggregated on a server for training, with a strong emphasis on feature engineering and algorithmic optimization to maximize detection metrics. A prevalent theme is the combination of sophisticated feature selection techniques with powerful classifiers. For instance, the authors of \cite{sun2024optimized} propose a method termed PSO-AdaBoost, which integrates Particle Swarm Optimization for feature selection with the AdaBoost classifier. Their evaluation on the NSL-KDD dataset \cite{dhanabal2015study} demonstrates performance superior to traditional algorithms like K-Nearest Neighbors (KNN) and Naive Bayes (NB). Similarly, the work in \cite{kilincer2023automated} utilizes Recursive Feature Elimination (RFE) with either Logistic Regression or an XGBoost Regressor to identify optimal features before employing a parameter-optimized Multi-layer Perceptron (MLP). This methodology is validated on a suite of IoMT datasets, including WUSTL-EHMS \cite{DTST22CONF}, ECU-IoHT \cite{ahmed2021ecu}, Intensive Care Unit (ICU) \cite{tao2018secured}, and ToN-IoT \cite{moustafa2021new}, with the XGB Regressor and MLP combination yielding particularly promising results.

To address the challenges of high-dimensional data and imbalanced classes, more complex deep learning architectures are proposed. The study in \cite{gu2023intrusion} introduces a framework using a Stacked Sparse Autoencoder (sSAE) to reduce dimensionality and computational memory requirements, followed by a Sliced Gated Recurrent Unit (SGRU) for classification on the AWID dataset \cite{kolias2015intrusion}. They address class imbalance with SMOTE and show their model outperforms numerous benchmarks, including DNN, Random Forest (RF), and LSTM hybrids, while also reducing model size. Further advancing architectural complexity, the authors of \cite{ravi2023deep} design a Multidimensional DL Model for smart healthcare enterprises. This model extracts features using separate CNN, bidirectional LSTM, and CNN-LSTM sub-models, concatenates these features, and uses fully connected layers for final classification. Evaluated on datasets including a KISTI enterprise network payload, KDDCup-99 \cite{kddcup99}, CICIDS2017 \cite{sharafaldin2018toward}, and WSN-DS \cite{almomani2016wsn}, AND UNSW-NB15 \cite{moustafa2016evaluation}, their solution shows particularly strong results on the enterprise dataset. Furthermore, an IDS designed for multi-cloud healthcare systems is proposed in \cite{gupta2022cybersecurity}. Their approach leverages a Deep Hierarchical Stacked Neural Network, reusing layers trained at the edge cloud level to create a pre-trained model at the core cloud. This method, tested on UNSW-BOT-IoT \cite{koroniotis2019towards} and UNSW-NB15 \cite{moustafa2016evaluation} datasets, is shown to enhance accuracy while reducing training time.

Research specifically targeting the IoMT domain often highlights the superiority of DL over traditional ML. The PSO-DNN model presented in \cite{chaganti2022particle}, tested on a dataset containing both network and medical features from WUSTL-EHMS \cite{DTST22CONF}, finds that DL models consistently outperform ML models like LR, KNN, DT, RF, and SVM. This advantage is attributed to DL's ability to handle complex, high-dimensional data. The critical importance of feature selection in this context is further explored in \cite{alalhareth2023improved}, where the LRGU-MIFS technique is introduced to select relevant features irrespective of data distribution, with optimal results found using a minimal feature set on the WUSTL-EHMS dataset. The PCA-GWO-DNN method in \cite{rm2020effective} also reports outperforming SVM, RF, NB, and KNN while reducing learning time. Beyond network features, the comparative study in \cite{hady2020intrusion} emphatically demonstrates that combining network and medical attributes significantly enhances the performance of algorithms like ANN, RF, KNN, and SVM compared to using either data type in isolation, underscoring the unique multi-modal nature of IoMT security.

While these centralized approaches achieve high performance, they possess fundamental limitations for the IoMT domain. First, they necessitate the collection of sensitive patient data to a central server, creating a significant privacy risk and a single point of failure. Second, the increasing complexity of these models, particularly DL architectures, renders them "black boxes," whose decisions are opaque and unexplainable. This lack of transparency is a major barrier to trust and adoption in healthcare, where understanding the rationale behind an alert is critical. Finally, as data volume grows, the scalability of constantly centralizing all data becomes a practical bottleneck.

To directly mitigate the privacy concerns of centralization, FL is introduced to enable collaborative model training without data ever leaving its source device. Several works have tailored FL for healthcare and IoMT environments. The framework proposed in \cite{al2023privacy} employs a differential privacy contractive deep autoencoder to fuse and privatize data from distributed cloud nodes before using a quantum DNN for detection. This solution achieves accuracy and detection rates exceeding 99\% on the WUSTL-EHMS and ICU datasets. The authors of \cite{otoum2021federated} advocate for federated transfer learning using DNN to enable collaborative training between edge and cloud models on the CICIDS2017 dataset \cite{sharafaldin2018toward}, showing their approach outperforms centralized benchmarks like SGD and Deep Belief Networks in generalization and incremental learning.

Addressing the specific security needs of critical infrastructure, the study in \cite{schneble2019attack} designs an FL-based IDS for Medical Cyber-Physical Systems (MCPS). Their method involves clustering historical data from the MIMIC dataset and implementing a dual-mode (learning/testing) protocol for end devices to minimize communication overhead during federated training. The system proves highly effective against simulated DoS, modification, and injection attacks. A different architectural approach, Hierarchical FL (HFL) based on hierarchical long-term memory, is put forward in \cite{singh2022dew}. This method constructs local models at dew-servers within a healthcare institution and aggregates them at the cloud level. Tested on the ToN-IoT and NSL-KDD datasets with PCA for dimensionality reduction, their HFL model surpasses the performance of LSTM, RNN, and GRU models.

FL effectively solves the data privacy problem inherent in centralized IDS by keeping data decentralized. However, the current research focus in FL-based IDS is almost exclusively on improving learning efficiency, accuracy, and reducing communication costs. A paramount shortcoming is that these solutions do not address the "black box" nature of the collaboratively trained global model. The decisions made by the global model remain just as opaque and unexplainable as those in a centralized system, failing to provide the interpretability needed for clinical trust and operational actionability.

The field of XAI for security is still emerging, with few applications in IoMT. A noteworthy instance is the work in \cite{tanveer2022xsru}, which proposes a bidirectional Simple Recurrent Unit (SRU) with skip connections for attack detection on the ToN-IoT dataset. While the model itself achieves high performance, the key contribution is the supplementary use of Local Interpretable Model-agnostic Explanations (LIME) to provide post-hoc, local explanations for individual predictions, offering insights into which features are most influential for a specific alert.

The application of XAI in \cite{tanveer2022xsru} represents a positive step but is indicative of the current limitations in the field. XAI is used as an external tool to explain a black-box model, providing only local rather than global interpretability. There is a lack of work on designing intrinsically interpretable models for IDS or on integrating XAI directly into the learning pipeline of a privacy-preserving framework like FL.

A systematic analysis of the reviewed literature identifies a persistent tripartite division within the field. Prevailing methodologies demonstrate proficiency in isolated domains yet fail to achieve a comprehensive synthesis. FL architectures \cite{otoum2021federated, schneble2019attack, singh2022dew} effectively address critical data privacy constraints through decentralized, collaborative model training; however, they inherently lack mechanisms for global model interpretability. Conversely, XAI techniques \cite{tanveer2022xsru} furnish post-hoc, local explainability for model predictions but are predominantly applied within centralized paradigms, neglecting integration with privacy-preserving infrastructures. This stands in contrast to conventional centralized ML/DL approaches \cite{al2023privacy, tanveer2022xsru, sun2024optimized, kilincer2023automated, gu2023intrusion, ravi2023deep, sharafaldin2018toward, chaganti2022particle, alalhareth2023improved, rm2020effective, hady2020intrusion}, which achieve high predictive performance at the expense of both data confidentiality and operational transparency. Consequently, a salient research gap persists: the absence of a unified architecture that concomitantly incorporates the privacy-by-design principles of FL, the transparent decision-making afforded by XAI, and the high analytical fidelity of advanced ML/DL models. This work is architected to address this deficit by introducing a novel IDS that synergistically amalgamates these three foundational components. The proposed framework is expressly designed to augment the security, privacy, and, fundamentally, the trustworthiness and ethical accountability of IoMT ecosystems.

\begin{table}
    \centering
    \begin{tabular}{|p{1cm}|p{5cm}|p{3cm}|p{2cm}|p{2cm}|p{2cm}|p{2cm}|} \hline 
         Ref &  Methods &  Datasets&  Learning approach&  use of the XAI&  Optimizing FL parameter \\ \hline 
         \cite{al2023privacy} &  differential privacy contractive deep autoencoder for data fusion and quantum DNN method for intrusion detection &  WUSTL-EHMS \cite{DTST22CONF} and ICU \cite{tao2018secured}&  central&  No&  No \\ \hline 
         \cite{otoum2021federated} &  DNN&  CICIDS2017 \cite{sharafaldin2018toward}&  FL&  No&  No \\ \hline 
         \cite{schneble2019attack} &  ANN&  MIMIC \cite{schneble2019attack}&  FL&  No&  Yes \\ \hline 
         \cite{singh2022dew} &  hierarchical long-term memory&  ToN-IoT and NSL-KDD \cite{dhanabal2015study}&  FL&  No&  No \\ \hline 
         \cite{tanveer2022xsru} &  bidirectional SRU with skip connections and LIME for XAI&  ToN-IoT \cite{moustafa2021new} &  central&  Yes&  No \\ \hline 
         \cite{sun2024optimized} &  PSO for features selections and AdaBoost for intrusion detection &  NSL-KDD \cite{dhanabal2015study} &  central&  No& No \\ \hline 
         \cite{kilincer2023automated} &  RFE for features selections then MLP for intrusion detection& WUSTL-EHMS \cite{DTST22CONF}, ECU-IoHT \cite{ahmed2021ecu}, ICU \cite{tao2018secured}, and ToN-IoT \cite{moustafa2021new} &  central&  No&  No \\ \hline 
         \cite{gu2023intrusion} & sSAE for dimensiality reduction then SGRU for intrusion detection &  AWID \cite{kolias2015intrusion}&  central&  No&  No \\ \hline 
         \cite{ravi2023deep} &  multidimensional DL model for features selection based on CNN, CNN-LSTM, and bidirectional LSTM then fully connected layers for intrusion detection&  KISTI, KDDCup-99 \cite{kddcup99}, CICIDS2017 \cite{sharafaldin2018toward}, WSN-DS \cite{almomani2016wsn} and UNSW-NB15 \cite{moustafa2016evaluation}&  central&  No&  No \\ \hline
         \cite{chaganti2022particle} &  PSO for features selections then DNN for intrusion detection&  WUSTL-EHMS \cite{DTST22CONF}&  central&  No&  No  \\ \hline
         \cite{alalhareth2023improved}&  integrates the LRGU technique  into the MIFS. for features selection then ML for intrusion detection&  WUSTL-EHMS \cite{DTST22CONF}&  central&  No&  No \\ \hline
         \cite{rm2020effective}&  PCA then GWO to minimize features then DNN for intrusion detection&  Kaggle intrusion data samples&  central&  No&  No \\ \hline
         \cite{kumar2021ensemble}&  ensemble learning that include NB, DT and RF&  ToN-IoT \cite{moustafa2021new}&  Fog-Cloud&  No&  No \\ \hline
         \cite{gupta2022cybersecurity}&  Deep Hierarchical Stacked Neural Networks &  UNSW-BOT-IoT \cite{koroniotis2019towards} and UNSW-NB15 \cite{moustafa2016evaluation}&  Multi-Cloud&  No&  No \\ \hline
         \cite{hady2020intrusion}&  ANN, RF, KNN, and SVM&  WUSTL-EHMS \cite{DTST22CONF}&  central&  No& No \\ \hline
    \end{tabular}
    \caption{Summary of research into anomaly detection in the context of the IoMT} \label{tbl1} 
    \label{tab:related work}
\end{table}

\section{METHODOLOGY FOR PROPOSED IDS MODEL} \label{METHODOLOGY FOR PROPOSED IDS MODEL}

In this section, the IoMT system under consideration for the proposed framework is explored, followed by an outline of the FL process, which involves local training at the end-device level and aggregation at the server level. Subsequently, the employed XAI method is detailed. The objective of this framework is to ensure that data collection, transfer, and processing are conducted securely, respecting their privacy and aligning with international standards in the medical domain.

\subsection{IoMT Systems}
The considered IoMT architecture is a comprehensive framework consisting of three distinct layers: the Data Acquisition Layer, the Personal Server Layer, and the Medical Server Layer, all operating within a client-server topology \cite{si2023survey}.

In layer 1, various types of medical devices equipped with sensors and actuators are employed to collect vital information and administer medications to patients that suffer from chronic diseases and also it can be used for fall detection for elderly persons or measuring the performance of athletes. These devices can be categorized into four types \cite{irfan2018internet, khan2012future}: implanted devices within the body, wearable devices, ambient devices capturing environmental data, and stationary devices found within hospitals. Given the energy limitations, wireless connections are established between these medical devices and mobile devices using low, or ultra-low-power wireless communication technologies such as Bluetooth, Zigbee, or NFC, thereby overcoming communication constraints \cite{zhang2015security}.

Moving to layer 2, the physiological data acquired by medical devices is transmitted to personal servers such as smartphones, laptops, or gateways \cite{sun2019security}. These servers remotely process, store, and enhance patient data by adding contextual information, compressing it, and encrypt it. After processing the data meticulously, it is sent to the hospital’s server using standardized formatting and long-range communication protocols like Wi-Fi, GSM or Ethernets \cite{hussain2019authentication}. This approach supports diverse communication capabilities and node mobility while enabling data resending in the event of network interruptions \cite{arya2020data, wazid2019iomt}.

\begin{figure*}[!ht]
    \centering
    \includegraphics[width=1\textwidth, height=0.7\textwidth]{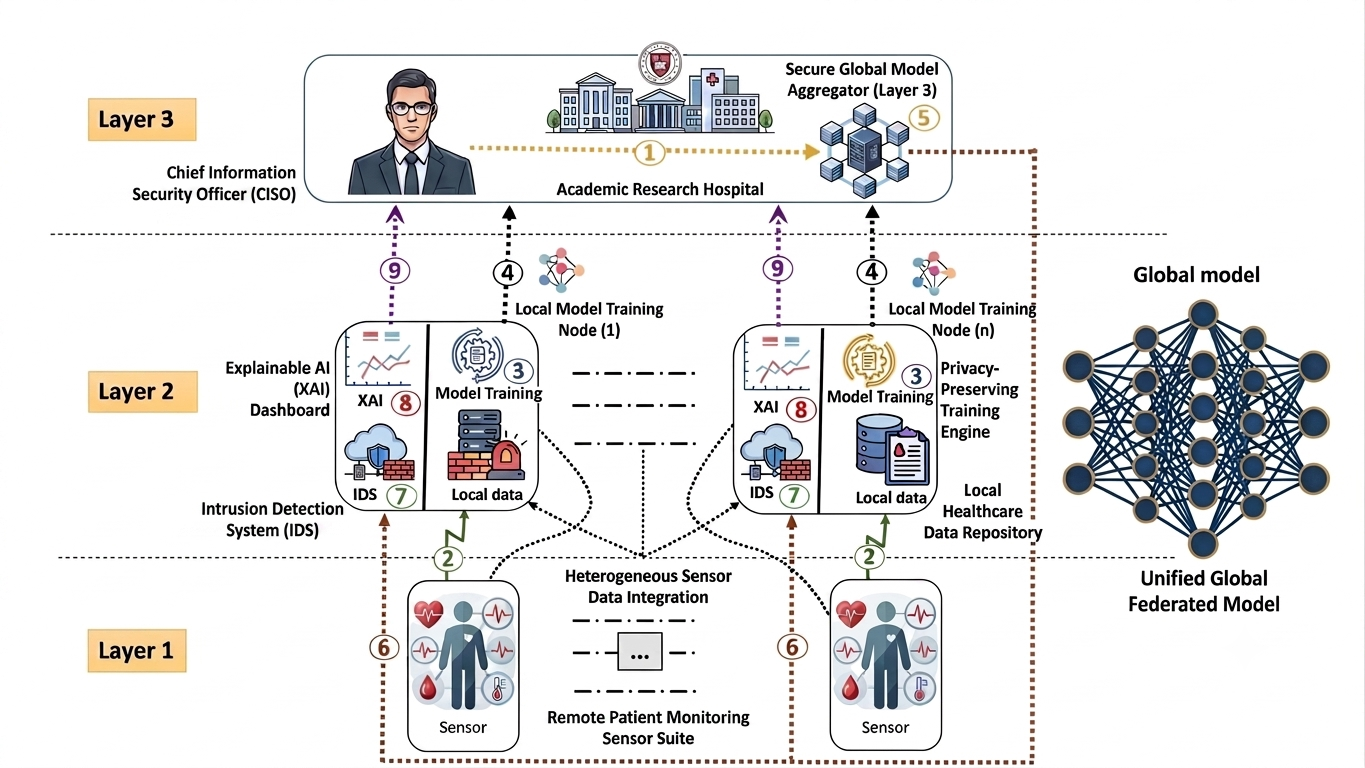}
    \caption{Comprehensive FL Architecture Implementation in IoMT}
    \label{FL_architecture}
\end{figure*}

Finally, in layer 3, the centralized medical server is responsible for handling messages transmitted from the mobile devices and relaying them back to the patients. The server must be equipped with high computational capacity to effectively handle incoming and outgoing communications as well as perform in-depth analysis of the received data using AI methods. Additionally, it features a cloud server for intelligent decision-making, aggregating and storing additional patient medical data. The collected data is accessible to doctors, patients, and the pharmacy department through an online interface or smartphone. Integration with Electronic Health Record (EHR) and Electronic Medical Record (EMR) systems ensures easy access to information and provides notifications for uploaded or received health data. The figure \ref{FL_architecture} provides an overview of the described system \cite{hathaliya2020exhaustive}.

\subsection{Local Training Process}
Utilizing DL  facilitates the detection of new and zero-day attacks, a capability lacking in signature-based methods. Furthermore, DL demonstrates the ability to identify complex patterns, thereby enhancing detection capabilities compared to traditional ML  approaches. Additionally, the significant feature selection process occurs automatically \cite{mclaughlin2017deep}.

In this context, the inclination is towards employing supervised learning in DL for IDS, leveraging ANN  for real-time anomaly detection. This architecture, simpler and less intricate than alternative models, enables adaptability and maintainability while delivering outstanding performance. Its capability to conduct real-time detection, as showcased in section \ref{EXPIREMENT SETUP AND RESULTS}, underscores its efficiency. Furthermore, the model’s straightforward design helps curb energy consumption within personal devices while contributing to reduced communication costs in the FL environment, which arise from exchanging model weights during the FL process instead of raw data.
There are three types of neural layer in an ANN: an input layer, one or more hidden layers and an output layer. Each neuron contains a threshold and connections to other neurons with weighted connections. Neurons get activated if their cumulative weight surpasses the threshold, transmitting signals to the subsequent layer \cite{ANN22CONF}.

The rectified linear unit (ReLU) introduced in \cite{glorot2011deep} serves as the activation function for the hidden layer, utilizing the MAX function (\ref{eq:relu}) to enable faster computation, prevent overfitting, and enhance overall model performance, while the He-Initialization method from \cite{he2015delving} is preferred for weight initialization due to its compatibility with ReLU (\ref{he_init}), ensuring efficient training dynamics. For the output layer, the sigmoid activation function (\ref{sigmoid}) is applied, making it well-suited for binary classification tasks. During model compilation, cross-entropy is employed as the loss function (\ref{crossEntropy}), and the Adam optimizer, described in \cite{kingma2014adam} (\ref{adam}), is chosen for its efficiency, combining the benefits of adagrad, which handles sparse gradients effectively, and RMSProp, which excels in optimizing non-stationary objectives, while its low memory requirements make it particularly suitable for large-scale datasets or models with a high number of parameters \cite{kingma2014adam}.

\begin{equation}\label{eq:relu} f(x) = \max(0, x) \end{equation}

\begin{equation}\label{he_init} \text{He-Initialization} = \mathcal{N}\left(0, \sqrt{\frac{2}{n}}\right) \end{equation}

Where:
\begin{align*} 
    & N : \text{Denotes the normal (Gaussian) distribution.} \\
    & \sqrt{\frac{2}{n}}: \text{Indicates the square root of the fraction} \frac{2}{n} \text{, where } n \text{ is the number of inputs in the layer.}
\end{align*}

\begin{equation} \label{sigmoid} \sigma(x) = \frac{1}{1 + e^{-x}} \end{equation}

Where:
\begin{align*} 
    & x : \text{represent the input to the sigmoid function. It can be the weighted sum of the inputs from the previous layer.}
\end{align*}

\begin{equation} \label{crossEntropy} 
H(y, p) = -\sum_{i} y_i \cdot \log(p_i) 
\end{equation}

Where:
\begin{align*} 
    & y_i : \text{is the i-th element of the true distribution, and $p_i$ is the i-th element of the predicted distribution.}
\end{align*}

\begin{equation} \label{adam}
    \begin{aligned} 
        m_t & = \beta_1 \cdot m_{t-1} + (1 - \beta_1) \cdot g_t \\
        v_t & = \beta_2 \cdot v_{t-1} + (1 - \beta_2) \cdot (g_t)^2 \\
        \hat{m}_t & = \frac{m_t}{1 - \beta_1^t} \\
        \hat{v}_t & = \frac{v_t}{1 - \beta_2^t} \\
        \theta_t & = \theta_{t-1} - \alpha \cdot \frac{\hat{m}_t}{\sqrt{\hat{v}_t} + \epsilon}
    \end{aligned}
\end{equation}

Where: 
\begin{align*} 
    & g_t: \text{ is the gradient.} \\
    & \alpha\ : \text{ is the learning rate.} \\ 
    & \beta_1  \beta_2: \text{ are the moment parameters.} \\
    & \epsilon: \text{ is a small constant to avoid division by zero.}
\end{align*}

\subsection{FL Porcess}

Improving the security of the IoMT  is crucial, and FL  emerges as a pivotal solution. FL facilitates model training by allowing the exchange of locally trained model weights among end-devices, avoiding the transmission of raw data. This innovative approach upholds data privacy standards, aligning seamlessly with regulations such as HIPAA and GDPR, and its decentralized nature mitigates the vulnerability of a centralized structure.

Beyond privacy preservation, FL reduces the volume of transmitted information by exchanging model parameters instead of raw datasets, thereby decreasing communication overhead and improving scalability in distributed healthcare infrastructures. These characteristics make FL particularly suitable for IoMT environments, where edge devices operate under constraints related to computational resources, energy availability, communication reliability, and heterogeneous network conditions. Accordingly, the proposed framework adopts a FL strategy that supports collaborative intrusion detection while preserving data confidentiality and enabling model interpretability through XAI.

Within the proposed framework, the Chief Information Security Officer (CISO) oversees the design, initialization, deployment, and maintenance of the AI model. The CISO ensures the random initialization of model weights and oversees its deployment at the server level. Additionally, the CISO is responsible for model maintenance. The hospital server, equipped with substantial computing and storage capabilities, orchestrates the development of the global model for all participating nodes. Responsibilities include registering personal devices, managing the global model, disseminating it, and selecting a subset of devices for FL participation, as outlined in Algorithm \ref{alg:cap}.

\begin{algorithm}
\caption{Hospital Server Update}\label{alg:cap} \Comment{run on server}
\begin{algorithmic}
\Require \textbf{Input}: Initial model weights $w_0$, number of communication rounds $R$, number of local epochs $E$, fraction of clients $Fr$, number of patients $M$, learning rate $\eta$.
\Ensure \textbf{Output}: Updated global model weights $w_{t+1}$ after aggregation.
\State \textbf{Description}: The hospital server initializes the global model with random weights using He-Initialization. It then iterates through a series of communication rounds, where a random subset of clients (personal devices) is selected to participate in training. Each client updates its local model using its own data and sends the updated weights back to the server. The server aggregates these local updates by averaging them to produce a new global model. This process repeats until the model converges or reaches the desired performance.

\State \textbf{Initialize model $w_0$ with He-Initialization} \Comment{Initialize global model weights using He-Initialization for efficient training.}
\State $R \gets$ Number of Round of communication \Comment{Set the total number of communication rounds between the server and clients.}
\State $E \gets$ Number of local epoch \Comment{Set the number of local training epochs for each client.}
\State $Fr \gets$ Fraction fit \Comment{Set the fraction of clients to be selected in each communication round.}
\State $M \gets$ Number of Patient \Comment{Set the total number of patients (clients) in the system.}
\State $\eta \gets$ learning rate \Comment{Set the learning rate for model updates.}

\For{$R = 1,2,3\dots$} \Comment{Begin communication rounds.}
\State $C \gets$ Random Set of $M \times Fr$ clients \Comment{Randomly select a subset of clients to participate in the current round.}

\For{patient $k \subset C$ in parallel} \Comment{Each selected client performs local training in parallel.}
\State $w_{(t+1)}^k \gets$ Patient Client Update $(k, w_t, E)$ \Comment{Client $k$ updates its local model using its data and sends the updated weights back to the server.}
\EndFor

\State $w_{(t+1)} \gets \sum_{k+1}^k \frac{\eta_k}{\eta} w_{(t+1)}^k$ \Comment{Aggregate the local model weights from all participating clients to update the global model.}
\EndFor
\end{algorithmic}
\end{algorithm}

\begin{algorithm}
\caption{Patient Client Update}\label{alg:cap2} \Comment{run on client}
\begin{algorithmic}
\Require \textbf{Input}: Current global model weights $w_t$, number of local epochs $E$, local data on the client device.
\Ensure \textbf{Output}: Updated local model weights $w_{t+1}^k$ after local training.
\State \textbf{Description}: Each selected client performs local training on its data for $E$ epochs using the Adam optimizer to minimize the binary cross-entropy loss function. The client updates its local model weights based on the training data and sends the updated weights back to the hospital server for aggregation.

\For{$e = 0$ to $e = E - 1$} \Comment{Perform local training for $E$ epochs.}
\State $w \gets$ use Adam optimizer to update $w$ to minimize the binary cross-entropy loss function \Comment{Update local model weights using the Adam optimizer.}
\EndFor

\State \textbf{return} $w$ to server \Comment{Send the updated local model weights back to the server.}
\end{algorithmic}
\end{algorithm}

\begin{figure}[!ht]
    \centering
    \includegraphics[
        width=\textwidth,
        height=0.95\textheight,
        keepaspectratio
    ]{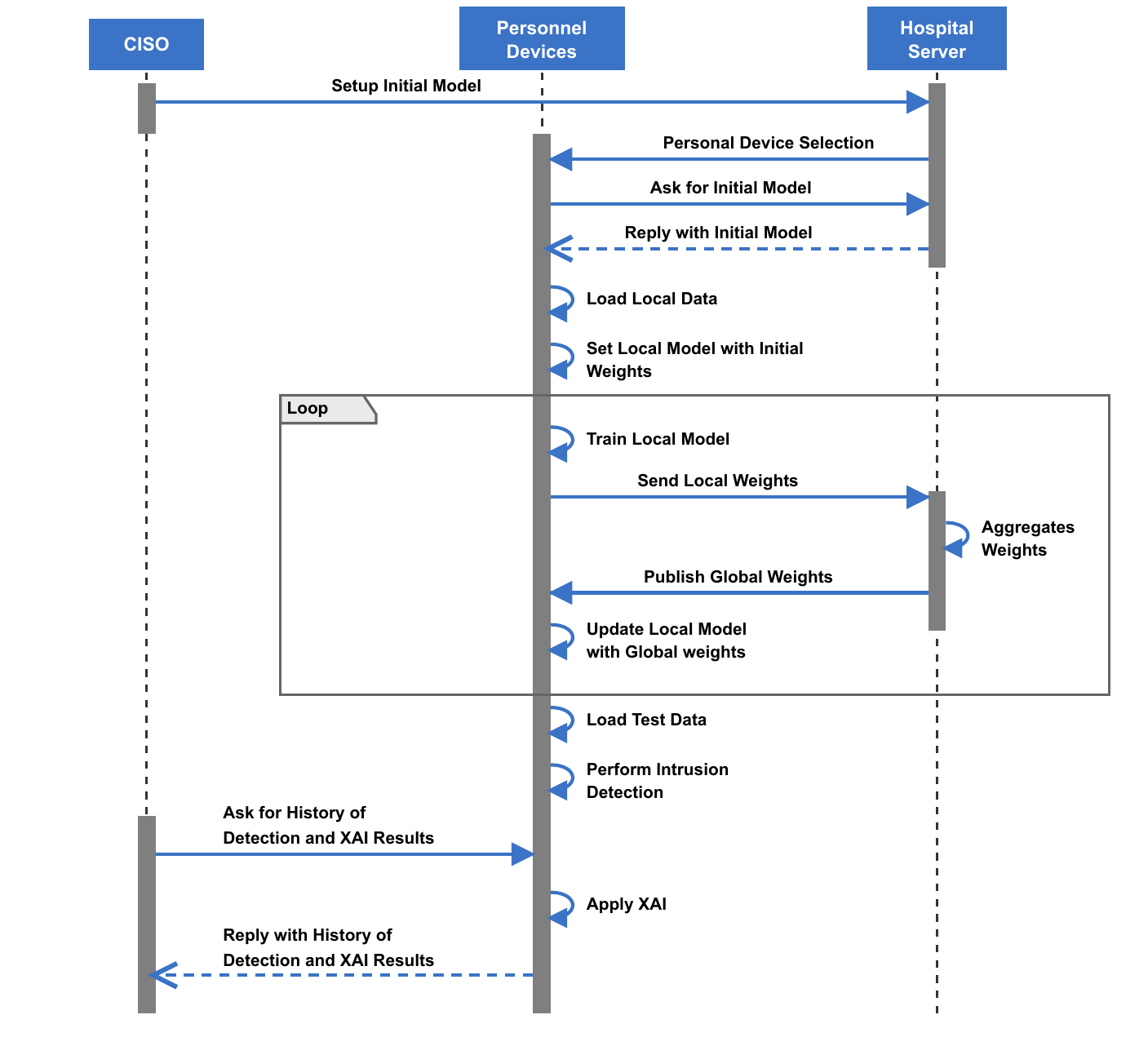}
    \caption{Flowchart depicting the operation of the proposed solutions.}
    \label{flowchart}
\end{figure}

Personal devices are represented by smartphones or equivalent edge devices equipped with sufficient storage capacity to retain data collected from associated medical sensors and adequate computational resources to perform local model training. These devices support both short-range and long-range wireless communications. Short-range communication technologies, such as Bluetooth Low Energy or Wi-Fi Direct, enable direct data acquisition from nearby medical sensors, whereas long-range communication technologies, including Wi-Fi and cellular networks (4G/5G), provide connectivity with the hospital server for exchanging model parameters. Consequently, personal devices act as intermediate edge nodes that bridge IoMT sensors and the centralized coordination server while preserving the locality of sensitive medical data. The overall workflow is illustrated in Figure \ref{FL_architecture} and detailed in the flowchart shown in Figure \ref{flowchart}. The proposed FL process consists of the following steps :

\begin{enumerate}
    \item \textbf{Initialization \& Model Download:} 
    The security administrator initializes the global model with random weights at the hospital server. Personal devices selected for FL download these initial weights to initialize their local models.

\item \textbf{IoMT Data Acquisition:} 
    Medical sensors send captured data to paired personal devices using short-range wireless communication (e.g., Bluetooth Low Energy), ensuring low-power data transfer.

\item \textbf{Local Training:} 
    Once personal devices have accumulated a sufficient amount of data, local models are trained for a predefined number of epochs, as outlined in Algorithm~\ref{alg:cap2}. This approach reduces communication costs by minimizing the number of communication rounds required for model convergence, thereby optimizing bandwidth usage.

\item \textbf{Secure Transmission of Updates:} 
    Updated local model weights are transmitted securely to the hospital server using encrypted communication protocols.

\item \textbf{Federated Averaging (FedAvg) Aggregation:} 
    The hospital server aggregates the local model parameters received from participating devices using the Federated Averaging (FedAvg) algorithm, as defined in Equation (\ref{fed_avg}). Because the availability of personal devices may vary across communication rounds owing to factors such as intermittent connectivity, mobility, energy limitations, or local resource constraints, each aggregation round incorporates only the model updates received from devices that successfully complete the current training cycle. The aggregated parameters are then used to construct an updated global model representing the collective knowledge learned across participating devices.

\item \textbf{Global Model Broadcast:} 
    The updated global model weights are sent back to the participating personal devices.

\item \textbf{Model Deployment \& Intrusion Detection:}  
    Steps 2 through 6 are repeated until the weight modifications become insignificant and the model converges. Once the global model achieves the desired performance, it is deployed for real-time intrusion detection across the IoMT network.

\item \textbf{Explanation Generation:} 
    Following deployment, the XAI process is applied to the final trained model. Using the The SHapley Additive exPlanations (SHAP) method, feature attribution values are calculated to provide a transparent, auditable record of the model's decision-making logic for each prediction.

 \item \textbf{Centralized Audit and Reporting:} 
    The outcomes derived from XAI analysis, along with historical intrusion detection data, are transmitted to the CISO either periodically or upon request, facilitating iterative refinement and debugging of the intrusion detection model, enhancing user trust through data accessibility on personal devices, and ensuring regulatory transparency and compliance with international healthcare standards.
\end{enumerate}

This FL process is well-suited to IoMT systems, which continuously generate data, enriching the models used for intrusion detection. The integrated use of FL and XAI effectively addresses the critical constraints of privacy, limited device resources, and the need for transparent decision-making in healthcare environments.

\begin{equation} \label{fed_avg} 
 W_{\text{new}} = \frac{1}{N} \sum_{i=1}^{N} w_i 
\end{equation}. 

where:
\begin{align*}
    W_{\text{new}} & : \text{The new global model after aggregation.} \\
    w_i & : \text{The local model update from client } i. \\
    N & : \text{The total number of participating clients.}
\end{align*}

\subsection{XAI Process}

ML has demonstrated its effectiveness in anomaly detection, particularly in identifying zero-day attacks and new vulnerabilities. This makes ML more advantageous compared to signature-based methods. However, non-transparent ML models, such as ANN, operate as black boxes, making it challenging to explain the reasoning behind the classification of instances, such as identifying an attack. This results in time consumption and ambiguity in analyzing predictions generated by these models.  XAI provides a solution by assisting model designers in determining the impact of each attribute on the classification process. This proves beneficial for model designers, aiding in debugging, enhancing the ML model, gaining insights into its decision-making process, and fostering increased trust in the model. 

The implementation of ANN involves deploying a non-transparent model, necessitating post-hoc explanation. Consequently, the feature relevance method is employed, as it aligns well with the framework. Among the various methodologies for determining feature relevance, SHAP stands out as a widely utilized and agnostic method, introduced by Lundberg and Lee \cite{lundberg2017unified}. Rooted in cooperative game theory, SHAP relies on Shapley values, offering both local and global scopes of explanation. Within the proposed framework, the global scope explanation is applied.

The fundamental principle of SHAP involves attributing a value that represents a median marginal contribution to the prediction across all possible feature combinations, calculated by comparing model performances with and without specific attributes. Mathematically, the Shapley value can be defined as follows (\ref{chap_value_formula}): 

 \begin{equation}  \label{chap_value_formula}
 \phi_i(v) = \sum_{S \subseteq N \setminus \{i\}} \frac{|S|!(|N|-|S|-1)!}{|N|!} [v(S \cup \{i\}) - v(S)] 
 \end{equation}

where:
\begin{align*}
    & \phi_i(v): \text{Shapley value for player (feature) } i \text{ in the cooperative game } v. \\
    & N: \text{The set of all players (features).} \\
    & S: \text{A coalition of players excluding player } i. \\
    & v(S \cup \{i\}): \text{The value of the coalition } S \text{ extended by adding player } i. \\
    & v(S): \text{The value of the coalition } S.
\end{align*}

The SHAP method presented in this framework serves to enhance the trust of model designers in the predictions made by their models. Additionally, it provides regulators with a means to verify the conformity of the framework with international standards concerning the explainability of decision-making, especially in critical sectors such as healthcare. However, for end-users lacking in-depth knowledge of AI and facing challenges in explaining and interpreting SHAP results, a historical record of predictions made by the framework is sufficient to demonstrate its effectiveness in intrusion detection. This, in turn, increases the trust of such users in the proposed framework.

\subsection{Ethical Considerations}
The proposed framework makes significant contributions to ethical considerations in healthcare by addressing core principles such as data protection, transparency, and accountability.

\begin{itemize}
\item \textbf {Data Protection and Privacy}: The framework leverages FL for IDS, ensuring that patient data remains localized on their devices. Only model weights are securely shared, minimizing privacy risks and ensuring compliance with stringent regulations like HIPAA and GDPR. This approach reduces the likelihood of data breaches and cyberattacks, thereby fostering greater trust in the system.

\item \textbf {Transparency in AI Decisions}: By integrating XAI methods such as SHAP, the framework provides clear and interpretable explanations for the model's decisions. This transparency enables stakeholders—including model designers, healthcare providers, and regulators—to understand how decisions are made, ensuring that AI outcomes are justifiable and aligned with user expectations and regulatory requirements.

\item \textbf {Accountability and Bias Mitigation}: The framework prioritizes fairness by utilizing diverse datasets and applying XAI techniques to detect and correct biases. This ensures that predictions are equitable and that any errors or biases in the model can be promptly identified and addressed. By doing so, it enhances accountability among system designers and operators, promoting responsible AI deployment.

\item \textbf {Compliance with International Standards}: The framework adheres to global standards such as ISO/IEC 27001 and ISO/IEC 27701, incorporating robust security measures like encrypted communication and decentralized data storage. Additionally, XAI ensures the traceability and transparency of model decisions, further reinforcing the framework’s credibility and reliability in a global context.
\end{itemize}

\section{EXPERIMENT SETUP AND RESULTS } \label{EXPIREMENT SETUP AND RESULTS}

In this section, evaluation of the proposed framework is conducted from various perspectives, utilizing a range of datasets \cite{ayoub1609_IDS_IoMT}. Initially, the impact of FL parameter modifications on performance is analyzed. Subsequently, the parameters yielding the best results are selected for comparison with the centralized approach. SHAP results are then employed to explain and interpret the outcomes of the proposed framework. Furthermore, a comprehensive discussion of the obtained results is presented, offering valuable insights and interpretations.

The practical challenges in FL, such as device heterogeneity, network latency, and connectivity, are critical for real-world deployment. To streamline the analysis, several simplifying assumptions are introduced. First, it is presumed that the IoMT devices involved in the FL process have sufficient computational capabilities to conduct local model training. While these devices may vary in processing power, memory, and energy constraints, they are assumed to meet the minimum requirements necessary for training local models. This assumption shifts the focus toward optimizing the FL process, excluding extreme cases of severely resource-constrained devices.

Second, the network latency between IoMT devices and the central server is assumed to remain within acceptable thresholds for real-time communication. This ensures timely interactions during the FL process.

Finally, stable connectivity is presumed throughout the FL training phase. Although real-world scenarios may experience occasional disruptions, the devices are assumed to stay connected long enough to complete local training and exchange updates with the server. This allows the analysis to concentrate on the FL process itself, without addressing frequent disconnections or network instability.

\subsection{Experimental Environment}
The experiments are conducted using Google Colab Pro as the experimental platform. This cloud-based environment, built upon Jupyter Notebook, provides GPU-enabled virtual machines that significantly accelerate the training of deep learning models. The availability of GPU resources also enables the concurrent simulation of multiple federated clients during FL experiments, thereby improving computational efficiency and reducing the execution time of large-scale simulations. The experimental environment operates under the Linux operating system and provides 12.7 GB of RAM and 166.8 GB of storage, which are sufficient to support data preprocessing, model training, and performance evaluation.

The implementation relies on several software libraries, including TensorFlow, Pandas, NumPy, scikit-learn, SHAP for XAI, and Flower as the FL framework. Flower is selected because it provides a flexible and scalable simulation framework specifically designed for FL. In particular, it supports the simulation of heterogeneous clients with diverse computational capabilities and communication characteristics, enabling realistic evaluation of distributed IoMT environments. Furthermore, Flower facilitates large-scale FL experimentation while allowing the seamless transition from simulation to deployment on real edge devices, thereby enhancing the reproducibility and extensibility of experimental studies  \cite{beutel2022flower}. Collectively, these software libraries provide the functionalities required for model development, federated optimization, explainability analysis, and comprehensive experimental evaluation.

\subsection{Dataset Description}

Experiments are carried out on four distinct datasets containing attacks that can potentially impact the availability, confidentiality, and integrity of IoMT systems. These datasets include NSL-KDD, UNSW-NB15, and ToN-IoT, all comprising network data, and WUSTL-EHMS, which encompasses both network and medical data.

\begin{figure}[!ht]
    \centering
    \includegraphics[
        width=\textwidth,
        height=0.9\textheight,
        keepaspectratio
    ]{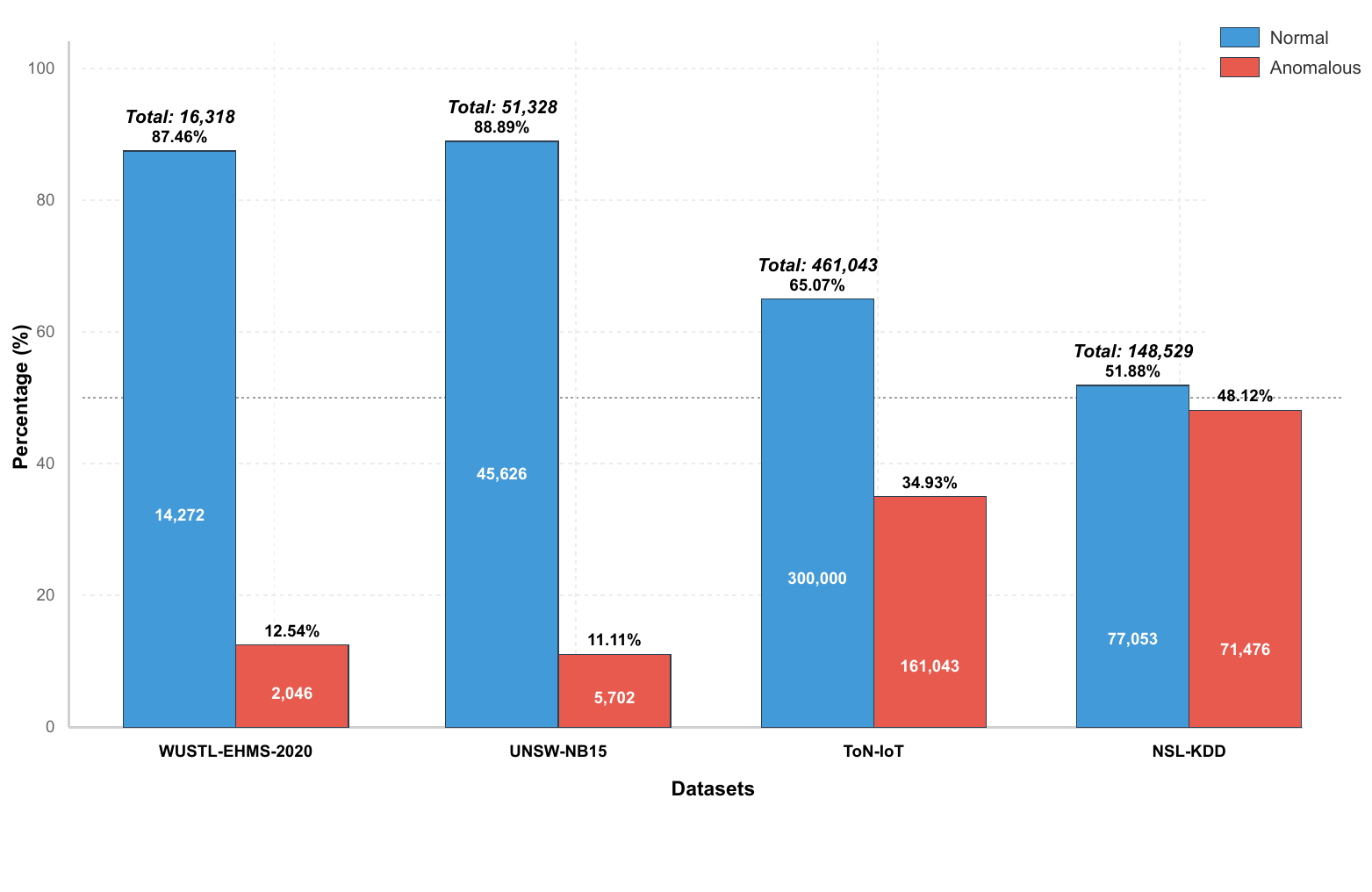}
    \caption{Class Distribution of Normal vs. Anomalous Instances Across Datasets}
    \label{Distribution}
\end{figure}

Each dataset exhibits unique characteristics in terms of attack diversity, feature representation, and data composition, as detailed in the subsequent descriptions. Figure \ref{Distribution} illustrates the comparative distribution of normal versus anomalous instances across all datasets, providing a comprehensive overview of their respective data distributions.

\begin{itemize}
\item \textbf {NSL-KDD : } The KDD99 and NSL-KDD datasets \cite{dhanabal2015study} are created by the IST division at the Lincoln Laboratories of the Massachusetts Institute of Technology. To generate the DARPA 98 dataset from raw packets, they develop a simulation testbed within the U.S. Air Force LAN system that includes both normal and attack scenario traffic. Subsequently, this dataset is renamed KDD99 and includes data characteristics derived from packets. The NSL-KDD dataset, an improved version of KDD99, is later developed to address limitations of the original dataset, such as removing redundant data and achieving a better balance between samples in the training and testing sets. The NSL-KDD dataset encompasses various attack categories, including Probing, Remote to Local (R2L), DoS, and User to Root (U2R), alongside a "Normal" class representing legitimate network traffic. Comprising 41 features, the dataset includes network connection attributes such as protocol type, service, source and destination IP addresses, and source and destination ports, among others.

\item \textbf {UNSW-NB15 : } The UNSW-NB15 dataset is officially released in 2015 by the Cyber Range Lab \cite{moustafa2016evaluation}, which operates under the auspices of the prestigious Australian Center for Cyber Security. Due to its remarkable utility, the dataset becomes a common choice for researchers within the cyber security domain, particularly among the research community affiliated with the Australian Centre for Cyber Security (ACCS). In the case of the UNSW-NB15 dataset, the authors opt to utilize unprocessed network packets, which are generated using the highly regarded IXIA Perfectstorm program. As part of the dataset evaluation, a comprehensive range of nine attack scenarios are meticulously implemented, encompassing diverse types such as DoS, fuzzes, analysis, backdoor, generic, reconnaissance, shellcode, exploits, and worms. To provide a holistic representation of network traffic, the dataset also includes a dedicated "Normal" class, specifically designed to capture legitimate network activity. Notably, a total of 49 network traffic features are meticulously extracted from the dataset, employing the robust Argus and Bro-IDS programs as essential analytical tools.

\item \textbf {ToN-IoT : } The ToN-IoT dataset \cite{moustafa2021new}, released by the IoT Lab of UNSW Canberra Cyber, addresses the limitations of existing datasets by collecting heterogeneous data from IoT and IIoT sources. It includes telemetry data, system logs, and system network traffic, providing a realistic representation of IoT networks. The dataset enables the evaluation of AI-based cybersecurity applications and features diverse attack scenarios such as XSS, DDoS, DoS, password cracking, reconnaissance, MITM, ransomware, backdoors, and injection attacks. Represented in CSV format, the dataset includes categorized columns for attack or normal behavior, facilitating analysis. The ToN-IoT dataset is a valuable resource for assessing the effectiveness of AI-enabled cybersecurity applications across IoT, network traffic, and operating systems.

\item \textbf {WUSTL-EHMS : }The WUSTL-EHMS dataset is generated using a real-time Enhanced Healthcare Monitoring System (EHMS) testbed \cite{DTST22CONF}. Due to the limited availability of a dataset that integrates these biometrics, this testbed accrues both the network flow metrics and patients' biometrics. This dataset comprises MITM attacks, spoofing and data injection. The spoofing attack merely sniffs the protocols passing through the gateway and the server, thus infringing upon the confidentiality of the patient's data. The data injection attack is employed to dynamically modify the packets, thereby infringing upon the integrity of the data.
\end{itemize}

\subsection{Data Preprocessing }
During the data preprocessing phase, a systematic approach is employed to optimize the dataset for FL analysis. The pipeline begins with label mapping to assign binary values, enabling clear classification of instances. Following this, feature removal is performed to eliminate columns containing single values, null values, or data labeling features that could introduce bias.

The preprocessing continues with feature-type-specific transformations: Boolean features are converted to binary format, categorical features undergo either ordinal encoding or one-hot encoding based on their nature, and numerical features are standardized using StandardScaler to optimize their scale and model compatibility.

The processed data is then divided through an 80/20 train-test split. The training portion (80\%) undergoes federated training data partitioning where it is distributed across multiple clients (Partition 1, Partition 2, ..., Partition N) for local model training. Simultaneously, the testing portion (20\%) is reserved for central server evaluation of the global model's performance.

\begin{figure}[!p]
    \centering
    \includegraphics[
        width=\textwidth,
        height=0.95\textheight,
        keepaspectratio
    ]{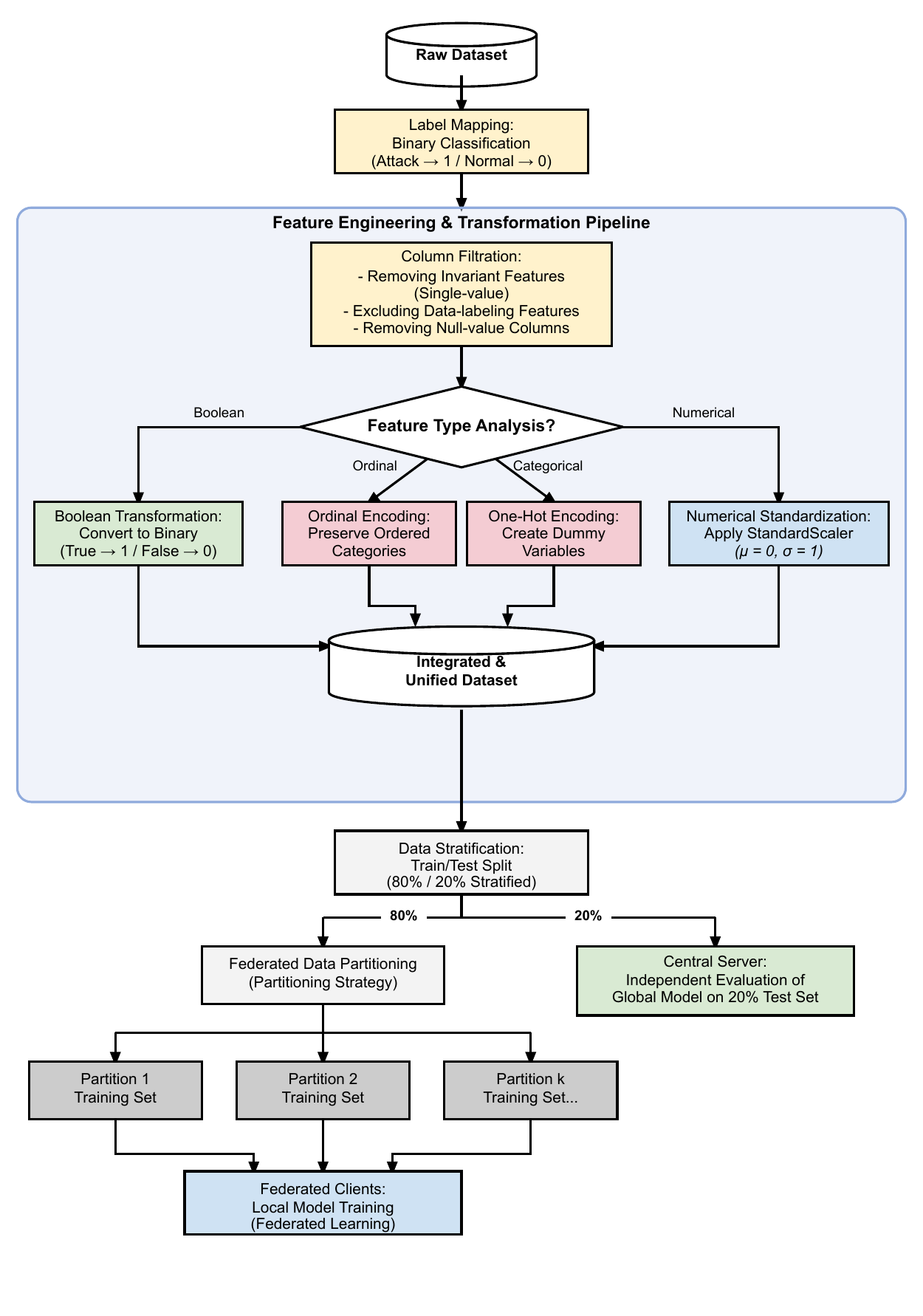}
    \caption{Data Preprocessing and Partitioning Pipeline for Federated Learning Systems}
    \label{Data Preparation}
\end{figure}

The flowchart depicted in Figure \ref{Data Preparation} outlines the comprehensive data preprocessing pipeline for FL systems, tracing the sequential transformation of raw data into partitioned datasets optimized for distributed model training and centralized evaluation.

\subsection{ML Algorithm}
The construction of various ANN models tailored to specific datasets involves distinctive configurations. For the UNSW-NB15 dataset, ANN are constructed with seven hidden layers, each comprising a different number of units: 150, 120, 90, 60, 30, 20, and 10 units, respectively. Conversely, the ToN-IoT dataset utilizes an ANN model with five hidden layers, incorporating 60, 40, 30, 20, and 10 units, respectively. In the case of the NSL-KDD dataset, a five-layered ANN model is established, featuring 80, 40, 30, 20, and 10 units. An ANN model is constructed for the WUSTL-EHMS dataset, comprising three hidden layers that are specified to contain 10, 20, and 40 units, respectively.

\subsection{Evaluation Metrics}
In the following formulas, True Positive (TP) denotes instances that are correctly identified as positive, False Negative (FN) represents instances that are incorrectly classified as negative, False Positive (FP) corresponds to instances that are incorrectly classified as positive, and True Negative (TN) refers to instances that are correctly classified as negative. These measures are indispensable for assessing the overall performance of a classification model \cite{wiki:confusionMatrix}. 

\begin{itemize}

\item \textbf{Confusion matrix : } The confusion matrix is a tabular representation that provides a summary of the performance of a classification model through the counts of TP, TN, FP, and FN \cite{wiki:confusionMatrix}.

\item \textbf{Recall : } is a performance metric that quantifies the accurate identification of positive instances by a model or system, relative to the overall count of positive instances. To calculate it, the quantity of TP is divided by the sum of TP and FN. Recall is mathematically represented as \cite{precision-and-recall} :

\begin{center}$Recall = \frac{TP}{TP + FN}$ \end{center}

\item \textbf{F1-score : } It enables the assessment of the accuracy of a model and integrates precision and recall metrics. A high F1 score signifies  small FP and FN \cite{FSCR22CONF} : 

\begin{center} $F1-score = 2 \times \frac{Precision \times Recall}{Precision + Recall}$ \end{center}

\item \textbf{Precision : } is a performance metric that quantifies the accuracy of positive identifications made by a model or system. It measures the ratio of TP to the sum of TP and FP. In other words, precision determines the proportion of positive identifications that are actually correct \cite{precision-and-recall}. 
Mathematically, precision can be represented as:

\begin{center}$Precision = \frac{TP}{TP + FP}$\end{center}

\item \textbf{Accuracy : } represents the proportion of the entire sample set that the model accurately predicts \cite{ACC22CONF}: 

\begin{center}
$Accuracy = \frac{TP + TN}{TP + TN + FP + FN}$ \end{center}

\item \textbf{Loss : } is a mathematical function used to quantify the disparity between the anticipated output of a ML model and the factual target value associated with a certain input \cite{wiki:lossFunction}. For binary classification tasks, this difference is typically measured using the binary cross-entropy loss function, as defined in Formula (\ref{crossEntropy}).

\item \textbf{AUC } (Area Under the Curve) : is a metric used to evaluate the performance of a binary classification model, specifically by measuring the area under the ROC (Receiver Operating Characteristic) curve. It quantifies the model's ability to distinguish between positive and negative classes, with values ranging from 0 to 1. An AUC of 1 indicates perfect classification, while 0.5 suggests no discriminative power, equivalent to random guessing  \cite{fawcett2006introduction}.

\end{itemize}

\subsection{Experiment Results }

The experimental study evaluates the proposed FL framework under operating conditions representative of IoMT environments. Unlike conventional distributed learning systems, IoMT infrastructures comprise heterogeneous personal devices interconnected with medical sensors, where computational resources, energy availability, communication reliability, and device participation may vary over time. Consequently, the configuration of the FL process plays a critical role in achieving efficient collaborative learning while accommodating the operational constraints inherent to healthcare systems.

Accordingly, the experimental analysis investigates three key FL parameters that are closely associated with these IoMT characteristics. The number of participating clients reflects the scalability of the collaborative learning process and the diversity of decentralized medical data contributing to the global model. The client participation fraction represents the dynamic availability of personal devices during each communication round, accounting for realistic situations in which devices may become temporarily unavailable because of mobility, intermittent wireless connectivity, battery limitations, or local resource constraints. The number of local training epochs determines the balance between local computation and global communication, directly influencing the communication frequency and the computational workload imposed on resource-constrained edge devices.


\newgeometry{margin=1cm} 

\begin{table*}[htbp]

   \begin{subtable}[b]{0.49\textwidth}

    \centering
    \begin{tabular}{|l|l|l|l|l|}
    \hline
        \textbf{metric / clients number} & \textbf{2} & \textbf{4} & \textbf{8} & \textbf{12} \\ \hline
        \textbf{Accuracy} & 0.9807 & 0.9808 & 0.9804 & 0.9792\\ \hline
        \textbf{Precision} & 0.9647 & 0.9618 & 0.9633 & 0.9624\\ \hline
        \textbf{Recall} & 0.9807 & 0.9841 & 0.9814 & 0.9787\\ \hline
        \textbf{F1-score} & 0.9726 & 0.9728 & 0.9723 & 0.9705\\ \hline
        \textbf{TP} & 31614 & 31724 & 31635 & 31614\\ \hline
        \textbf{TN} & 58816 & 58712 & 58769 & 58637\\ \hline
        \textbf{FP} & 1157 & 1261 & 1204 & 1303\\ \hline
        \textbf{FN} & 622 & 512 & 601 & 622\\ \hline
        \textbf{Loss} & 0.0521 & 0.0543 & 0.0524 & 0.0576 \\ \hline
        \textbf{AUC} & 0.9978 & 0.9977 & 0.9976 & 0.9967\\ \hline
        \textbf{communication rounds } & 52 & 56 &  48 & \cellcolor[HTML]{00FF00} 47 \\ \hline
        
    \end{tabular}
    \caption{ToN-IoT dataset.}
    \label{NumberClientsToN-IoT}

\end{subtable}
\hfill
\begin{subtable}[b]{0.5\textwidth}

    \centering  
    \begin{tabular}{|l|l|l|l|l|}
    \hline
        \textbf{metric / clients number} & \textbf{2} & \textbf{4} & \textbf{8} & \textbf{12} \\ \hline
        \textbf{Accuracy} & 0.9881 & 0.9883 & 0.9881 & 0.9884\\ \hline
        \textbf{Precision} & 0.9431 & 0.9424 & 0.9511 & 0.9511\\ \hline
        \textbf{Recall} & 0.9482 & 0.9508 & 0.9392 &  0.9398\\ \hline
        \textbf{F1-score} & 0.9456 & 0.9466 & 0.9451 &  0.9454\\ \hline
        \textbf{TP} & 1061 & 1064 & 1051 &  1031\\ \hline
        \textbf{TN} & 9083 & 9082 & 9093 &  9116\\ \hline
        \textbf{FP} & 64 & 65 & 54 &  53\\ \hline
        \textbf{FN} & 58 & 55 & 68 &  66\\ \hline
        \textbf{Loss} & 0.1009 & 0.0346 & 0.0542 &  0.0266\\ \hline
        \textbf{AUC} & 0.9982 & 0.9984 & 0.9983 & 0.9989\\ \hline
        \textbf{communication rounds } & 80 &  33 & 37 &  \cellcolor[HTML]{00FF00} 9\\ \hline
        
    \end{tabular}
    \caption{UNSW\_NB15 dataset}
    \label{NumberClientsUNSW_NB15}

\end{subtable}
\vspace{1cm} 
\begin{subtable}[b]{0.5\textwidth}

    \centering
    \begin{tabular}{|l|l|l|l|l|}
    \hline
        \textbf{metric / clients number} & \textbf{ 2} & \textbf{ 4} & \textbf{ 8} & \textbf{ 12} \\ \hline
        \textbf{Accuracy} & 0.9856 & 0.9904 & 0.9905 & 0.9904\\ \hline
        \textbf{Precision} & 0.9883 & 0.99 & 0.9904 & 0.9927 \\ \hline
        \textbf{Recall} & 0.9817 & 0.99 & 0.9898 & 0.9873\\ \hline
        \textbf{F1-score} & 0.985 & 0.99 & 0.99 & 0.99\\ \hline
        \textbf{TP} & 13990 & 14108 & 14106 & 14036\\ \hline
        \textbf{TN} & 15286 & 15309 & 15315 & 15383\\ \hline
        \textbf{FP} & 166 & 143 & 137 & 103 \\ \hline
        \textbf{FN} & 261 & 143 & 145 & 181\\ \hline
        \textbf{Loss} & 0.237 & 0.0822 & 0.0652 & 0.0632\\ \hline
        \textbf{AUC} & 0.9846 & 0.9958 & 0.9973 & 0.9971\\ \hline
        \textbf{communication rounds } & 50 & 27 & 18 & \cellcolor[HTML]{00FF00} 17 \\ \hline
        
    \end{tabular}    
    \caption{NSL-KDD dataset.}
    \label{NumberClientsNSL-KDD}

\end{subtable}
\hfill
\begin{subtable}[b]{0.5\textwidth}

    \centering  
    \begin{tabular}{|l|l|l|l|l|}
    \hline
        \textbf{metric / clients number} & \textbf{2} & \textbf{4} & \textbf{8} & \textbf{12} \\ \hline
        \textbf{Accuracy} & 0.9378 & 0.9381 & 0.9381 & 0.9381\\ \hline
        \textbf{Precision} & 0.9339 & 0.9381 & 0.9626 & 0.9157\\ \hline
        \textbf{Recall} & 0.53 & 0.53 & 0.515 & 0.5704\\ \hline
        \textbf{F1-score} & 0.6762 & 0.6773 & 0.671 & 0.7029\\ \hline
        \textbf{TP} & 212 & 212 & 206 & 239\\ \hline
        \textbf{TN} & 2849 & 2850 & 2856 & 2823\\ \hline
        \textbf{FP} & 15 & 14 & 8 & 22\\ \hline
        \textbf{FN} & 188 & 188 & 194 & 180\\ \hline
        \textbf{Loss} & 0.1973 & 0.2 & 0.1904 & 0.2061\\ \hline
        \textbf{AUC} & 0.8927 & 0.8996 & 0.8997 & 0.9166\\ \hline
        \textbf{communication rounds } & \cellcolor[HTML]{00FF00} 15 & 16 & 24 & 37\\ \hline
        
    \end{tabular}  
    \caption{WUSTL-EHMS dataset.}
    \label{NumberClientsWUSTL-EHMS}

\end{subtable}
\caption{Results of the Number of Clients Test on FL Performances Using Different Datasets}
\end{table*}

\begin{figure}[!ht]
    \centering
    \includegraphics[
        width=\textwidth,
        height=0.36\textheight,
        keepaspectratio
    ]{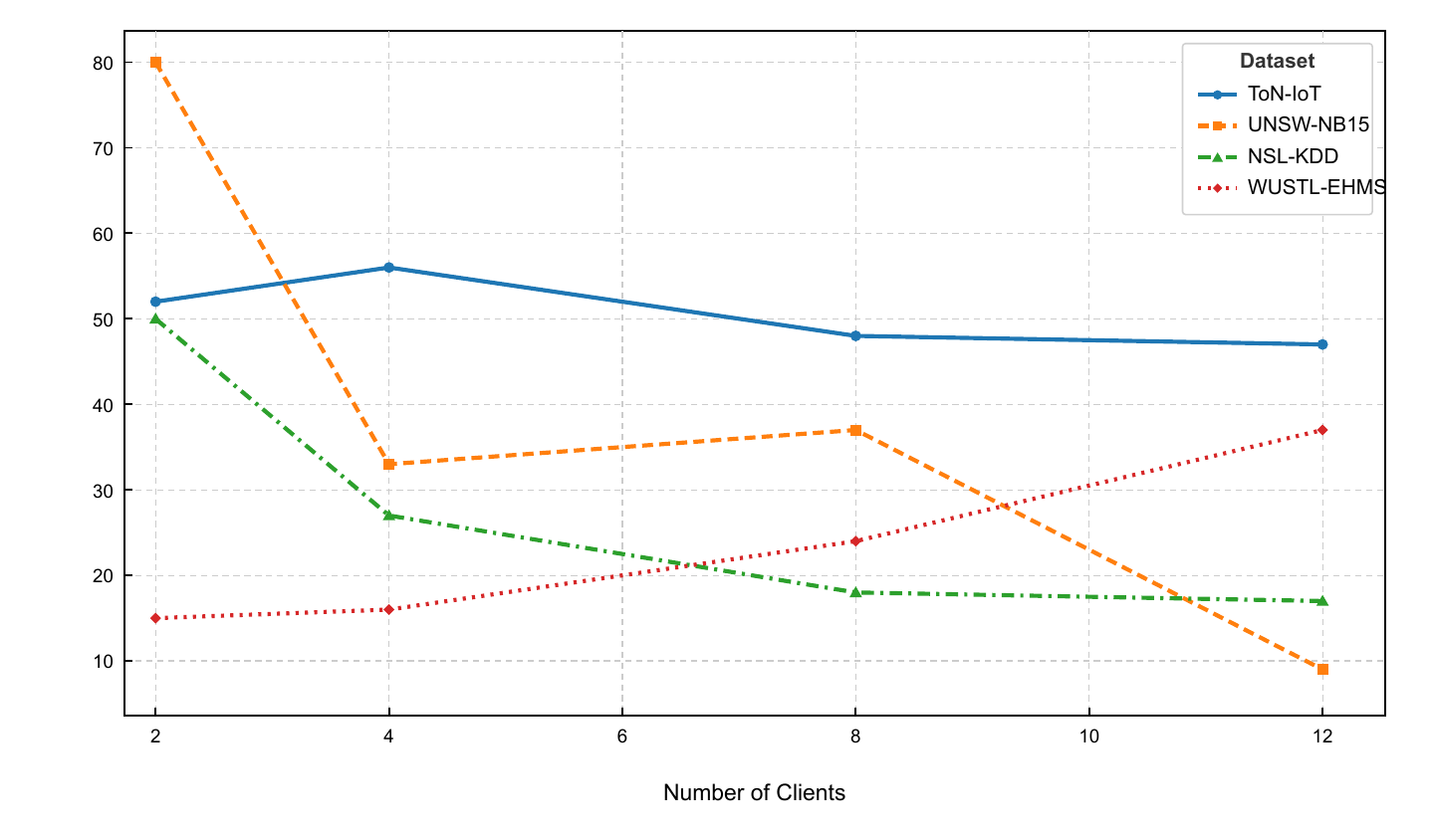}
    \caption{Impact of Client Pool Size on Federated Learning Convergence Using Different Datasets}
    \label{Client Number}
\end{figure}

\restoregeometry

\newgeometry{margin=1cm} 

\begin{table*}[htbp]
\begin{subtable}[b]{0.49\textwidth}
    
    \centering
    \begin{tabular}{|l|l|l|l|}
    \hline
        \textbf{metric / clients fractions} & \textbf{ 0.1} & \textbf{0.5} & \textbf{1} \\ \hline
        \textbf{Accuracy} & 0.9813 & 0.9805 & 0.9804 \\ \hline
        \textbf{Precision} & 0.9621 & 0.9622 & 0.9633 \\ \hline
        \textbf{Recall} & 0.9853 & 0.9828 & 0.9814 \\ \hline
        \textbf{F1-score} & 0.9735 & 0.9622 & 0.9723 \\ \hline
        \textbf{TP} & 31761 & 31683 & 31635 \\ \hline
        \textbf{TN} & 58721 & 58729 & 58769 \\ \hline
        \textbf{FP} & 1252 & 1244 & 1204 \\ \hline
        \textbf{FN} & 475 & 553 & 601 \\ \hline
        \textbf{Loss} & 0.049 & 0.0507 & 0.0524 \\ \hline
        \textbf{AUC} & 0.998 & 0.9979 & 0.9976 \\ \hline
        \textbf{communication rounds } & 52 & 60 & \cellcolor[HTML]{00FF00} 48 \\ \hline
        
    \end{tabular}
    \caption{Ton\_Iot dataset.}
    \label{FractionFitton_iot}

\end{subtable}
\hfill
\begin{subtable}[b]{0.5\textwidth}

    \centering 
    \begin{tabular}{|l|l|l|l|}
    \hline
        \textbf{metric / clients fractions} & \textbf{0.1} & \textbf{0.5} & \textbf{1} \\ \hline
        \textbf{Accuracy} & 0.9892 & 0.9884 & 0.9884 \\ \hline
        \textbf{Precision} & 0.9482 & 0.9587 & 0.9511 \\ \hline
        \textbf{Recall} & 0.9508 & 0.9316 & 0.9398 \\ \hline
        \textbf{F1-score} & 0.9495 & 0.945 & 0.9454 \\ \hline
        \textbf{TP} & 1043 & 1022 & 1031 \\ \hline
        \textbf{TN} & 9112 & 9125 & 9116 \\ \hline
        \textbf{FP} & 57 & 44 & 53 \\ \hline
        \textbf{FN} & 54 & 75 & 66 \\ \hline
        \textbf{Loss} & 0.0274 & 0.208 & 0.0266 \\ \hline
        \textbf{AUC} & 0.9989 & 0.999 & 0.9989 \\ \hline
        \textbf{communication rounds } & 12 & \cellcolor[HTML]{00FF00} 8 &  9 \\ \hline
        
    \end{tabular}  
     \caption{UNSW\_NB15 dataset.}
     \label{FractionFitUNSW_NB15}

\end{subtable}
\vspace{1cm} 
\begin{subtable}[b]{0.5\textwidth}

    \centering    
    \begin{tabular}{|l|l|l|l|}
    \hline
        \textbf{metric / clients fractions} & \textbf{ 0.1} & \textbf{ 0.5} & \textbf{ 1} \\ \hline
        \textbf{Accuracy} & 0.9863 & 0.9863 & 0.9905 \\ \hline
        \textbf{Precision} & 0.9891 & 0.9881 & 0.9904 \\ \hline
        \textbf{Recall} & 0.9823 & 0.9834 & 0.9898 \\ \hline
        \textbf{F1-score} & 0.9857 & 0.9857 & 0.99 \\ \hline
        \textbf{TP} & 13999 & 14014 & 14106 \\ \hline
        \textbf{TN} & 15298 & 15283 & 15315 \\ \hline
        \textbf{FP} & 154 & 169 & 137 \\ \hline
        \textbf{FN} & 252 & 237 & 145 \\ \hline
        \textbf{Loss} & 0.1766 & 0.2055 & 0.0652 \\ \hline
        \textbf{AUC} & 0.9888 & 0.9869 & 0.9973 \\ \hline
        \textbf{communication rounds } & 50 & 50 & \cellcolor[HTML]{00FF00} 18 \\ \hline
        
    \end{tabular}
    \caption{NSL-KDD dataset.}
    \label{FractionFitNSL-KDD}

\end{subtable}
\hfill
\begin{subtable}[b]{0.5\textwidth}

    \centering  
    \begin{tabular}{|l|l|l|l|}
    \hline
        \textbf{metric / clients fractions} & \textbf{ 0.1} & \textbf{ 0.5} & \textbf{ 1} \\ \hline
        \textbf{Accuracy} & 0.9384 & 0.939 & 0.9381 \\ \hline
        \textbf{Precision} & 0.9061 & 0.9313 & 0.9626 \\ \hline
        \textbf{Recall} & 0.555 & 0.5425 & 0.515 \\ \hline
        \textbf{F1-score} & 0.6884 & 0.6856 & 0.671 \\ \hline
        \textbf{TP} & 222 & 217 & 206 \\ \hline
        \textbf{TN} & 2841 & 2848 & 2856 \\ \hline
        \textbf{FP} & 23 & 16 & 8 \\ \hline
        \textbf{FN} & 178 & 183 & 194 \\ \hline
        \textbf{Loss} & 0.1807 & 0.2107 & 0.1904 \\ \hline
        \textbf{AUC} & 0.917 & 0.8956 & 0.8997 \\ \hline
        \textbf{communication rounds } & 30 & \cellcolor[HTML]{00FF00} 17 & 24 \\ \hline
        
    \end{tabular}
     \caption{WUSTL-EHMS dataset.}
     \label{FractionFitWUSTL-EHMS}

\end{subtable}
\caption{Results of the Fractions of Clients Test on FL Performances Using Different Datasets}
\end{table*}

\begin{figure}[!ht]
    \centering
    \includegraphics[
        width=\textwidth,
        height=0.36\textheight,
        keepaspectratio
    ]{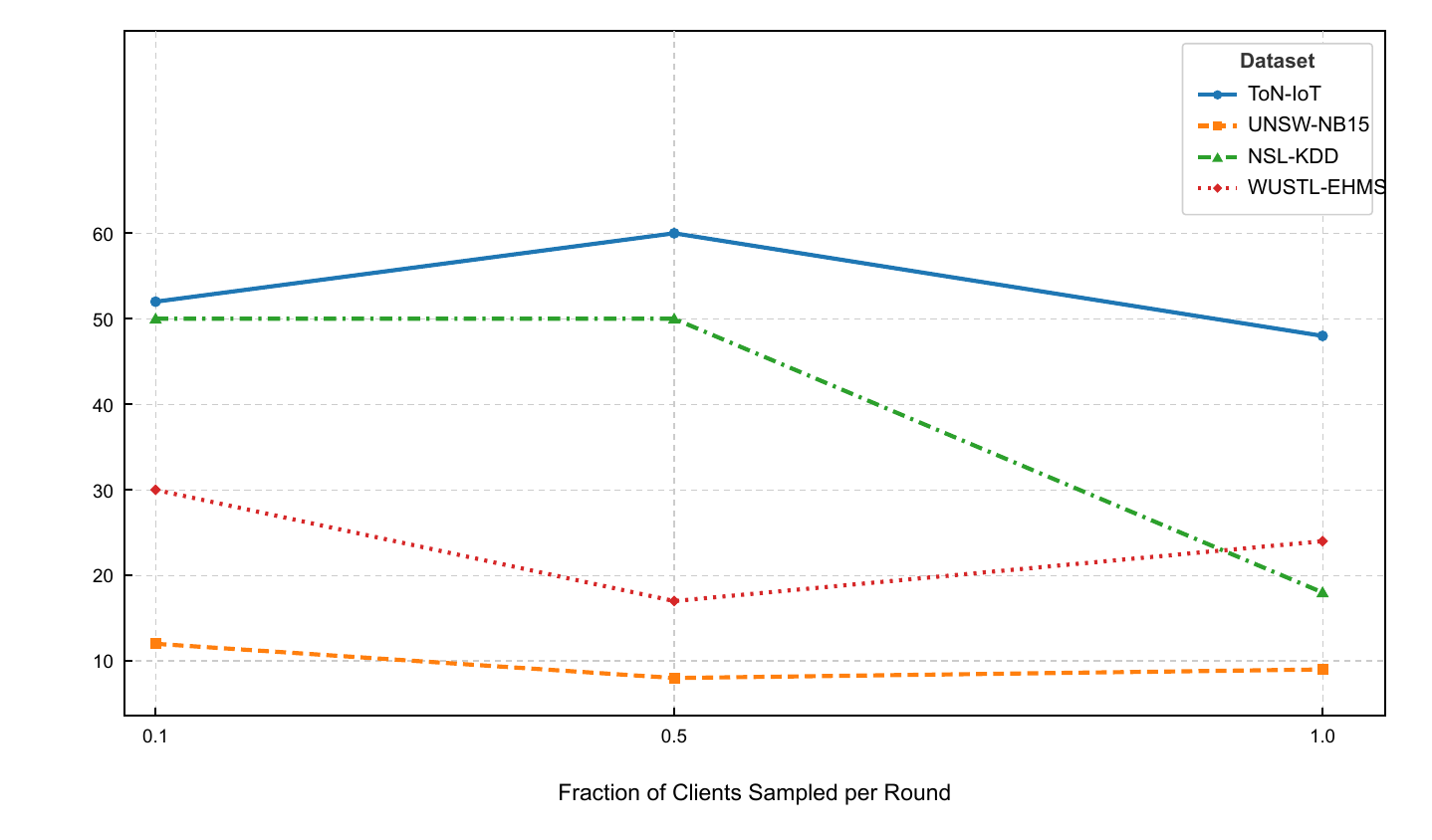}
    \caption{Effect of Client Participation Rate on Federated Learning Convergence Using Different Datasets}
    \label{Fraction}
\end{figure}

\restoregeometry

\newgeometry{margin=1cm} 

\begin{table*}[htbp]
\begin{subtable}[b]{0.49\textwidth}

    \centering    
    \begin{tabular}{|l|l|l|l|l|}
    \hline
        \textbf{metric / local epochs number} & \textbf{1} & \textbf{2} & \textbf{5} & \textbf{8} \\ \hline
        \textbf{Accuracy} & 0.98 & 0.9806 & 0.9809 & 0.9801 \\ \hline
        \textbf{Precision} & 0.961 & 0.9628 & 0.9634 & 0.9625 \\ \hline
        \textbf{Recall} & 0.9827 & 0.9825 & 0.9828 & 0.9814 \\ \hline
        \textbf{F1-score} & 0.97 & 0.973 & 0.973 & 0.972 \\ \hline
        \textbf{TP} & 31678 & 31671 & 31680 & 31636 \\ \hline
        \textbf{TN} & 58689 & 58749 & 58770 & 58740 \\ \hline
        \textbf{FP} & 1284 & 1224 & 1203 & 1233 \\ \hline
        \textbf{FN} & 558 & 565 & 556 & 600 \\ \hline
        \textbf{Loss} & 0.0538 & 0.0538 & 0.0497 & 0.0546 \\ \hline
        \textbf{AUC} & 0.9977 & 0.9977 & 0.9979 & 0.9977 \\ \hline
        \textbf{communication rounds } & 35 & 15 & 11 & \cellcolor[HTML]{00FF00} 6 \\  
        \hline
    \end{tabular}
    \caption{ton\_iot dataset.}
    \label{LocalEpochston_iot}

\end{subtable}
\hfill
\begin{subtable}[b]{0.5\textwidth}

    \centering  
    \begin{tabular}{|l|l|l|l|l|}
    \hline
        \textbf{metric / local epochs number} & \textbf{1} & \textbf{2} & \textbf{5} & \textbf{8} \\ \hline
        \textbf{Accuracy} & 0.9882 & 0.9881 & 0.989 & 0.9888 \\ \hline
        \textbf{Precision} & 0.9516 & 0.943 & 0.9408 & 0.948\\ \hline
        \textbf{Recall} & 0.9425 & 0.9512 & 0.9556 & 0.9521 \\ \hline
        \textbf{F1-score} & 0.9470 & 0.9471 & 0.9481 & 0.95 \\ \hline
        \textbf{TP} & 1082 & 1092 & 1097 & 1093 \\ \hline
        \textbf{TN} & 9063 & 9052 & 9049 & 9058 \\ \hline
        \textbf{FP} & 55 & 66 & 69 & 60 \\ \hline
        \textbf{FN} & 66 & 56 & 51 & 55 \\ \hline
        \textbf{Loss} & 0.1047 & 0.0476 & 0.0301 & 0.0341 \\ \hline
        \textbf{AUC} & 0.998 & 0.998 & 0.9985 & 0.9984 \\ \hline
        \textbf{communication rounds } & 25 & 10 & 5 & \cellcolor[HTML]{00FF00} 3 \\ 
        \hline
        
    \end{tabular}    
    \caption{UNSW\_NB15 dataset.}
    \label{LocalEpochsUNSW_NB15}

\end{subtable}
\vspace{1cm} 
\begin{subtable}[b]{0.5\textwidth}

    \centering    
    \begin{tabular}{|l|l|l|l|l|}
    \hline
        \textbf{metric / local epochs number} & \textbf{1} & \textbf{2} & \textbf{5} & \textbf{8} \\ \hline
        \textbf{Accuracy} & 0.9869 & 0.9858 & 0.9865 & 0.9866 \\ \hline
        \textbf{Precision} & 0.993 & 0.9925 & 0.9953 & 0.9899 \\ \hline
        \textbf{Recall} & 0.9793 & 0.9775 & 0.9762 & 0.9818 \\ \hline
        \textbf{F1-score} & 0.9861 & 0.985 & 0.986 & 0.9859 \\ \hline
        \textbf{TP} & 13841 & 13815 & 13797 & 13876 \\ \hline
        \textbf{TN} & 15472 & 15465 & 15505 & 15429 \\ \hline
        \textbf{FP} & 98 & 105 & 65 & 141 \\ \hline
        \textbf{FN} & 292 & 318 & 336 & 257 \\ \hline
        \textbf{Loss} & 0.0704 & 0.0726 & 0.0598 & 0.1652 \\ \hline
        \textbf{AUC} & 0.997 & 0.9971 & 0.9974 & 0.9897 \\ \hline
        \textbf{communication rounds } & 14 & \cellcolor[HTML]{00FF00} 7 & 12 & 10 \\ \hline
        
    \end{tabular}   
    \caption{NSL-KDD dataset.}
    \label{LocalEpochsNSL-KDD}

\end{subtable}
\hfill
\begin{subtable}[b]{0.5\textwidth}

    \centering   
    \begin{tabular}{|l|l|l|l|l|}
    \hline
        \textbf{metric / local epochs number} & \textbf{1} & \textbf{2} & \textbf{5} & \textbf{8} \\ \hline
        \textbf{Accuracy} & 0.9308 & 0.9286 & 0.9286 & 0.9326 \\ \hline
        \textbf{Precision} & 0.918 & 0.9367 & 0.96 & 0.9074 \\ \hline
        \textbf{Recall} & 0.5341 & 0.5045 & 0.4909 & 0.5568 \\ \hline
        \textbf{F1-score} & 0.6753 & 0.6558 & 0.65 & 0.69 \\ \hline
        \textbf{TP} & 235 & 222 & 216 & 245 \\ \hline
        \textbf{TN} & 2803 & 2809 & 2815 & 2799 \\ \hline
        \textbf{FP} & 21 & 15 & 9 & 25 \\ \hline
        \textbf{FN} & 205 & 218 & 224 & 195 \\ \hline
        \textbf{Loss} & 0.2222 & 0.2219 & 0.2247 & 0.2098 \\ \hline
        \textbf{AUC} & 0.8841 & 0.8766 & 0.8799 & 0.901 \\ \hline
        \textbf{communication rounds } & 18 & 7 & 3 & \cellcolor[HTML]{00FF00} 3 \\ \hline
        
    \end{tabular}   
\caption{WUSTL-EHMS dataset.}\label{LocalEpochsWUSTL-EHMS}

\end{subtable}
\caption{Results of the Number of Local Epochs Test on FL Performances Using Different Datasets}
\end{table*}

\begin{figure}[!ht]
    \centering
    \includegraphics[
        width=\textwidth,
        height=0.36\textheight,
        keepaspectratio
    ]{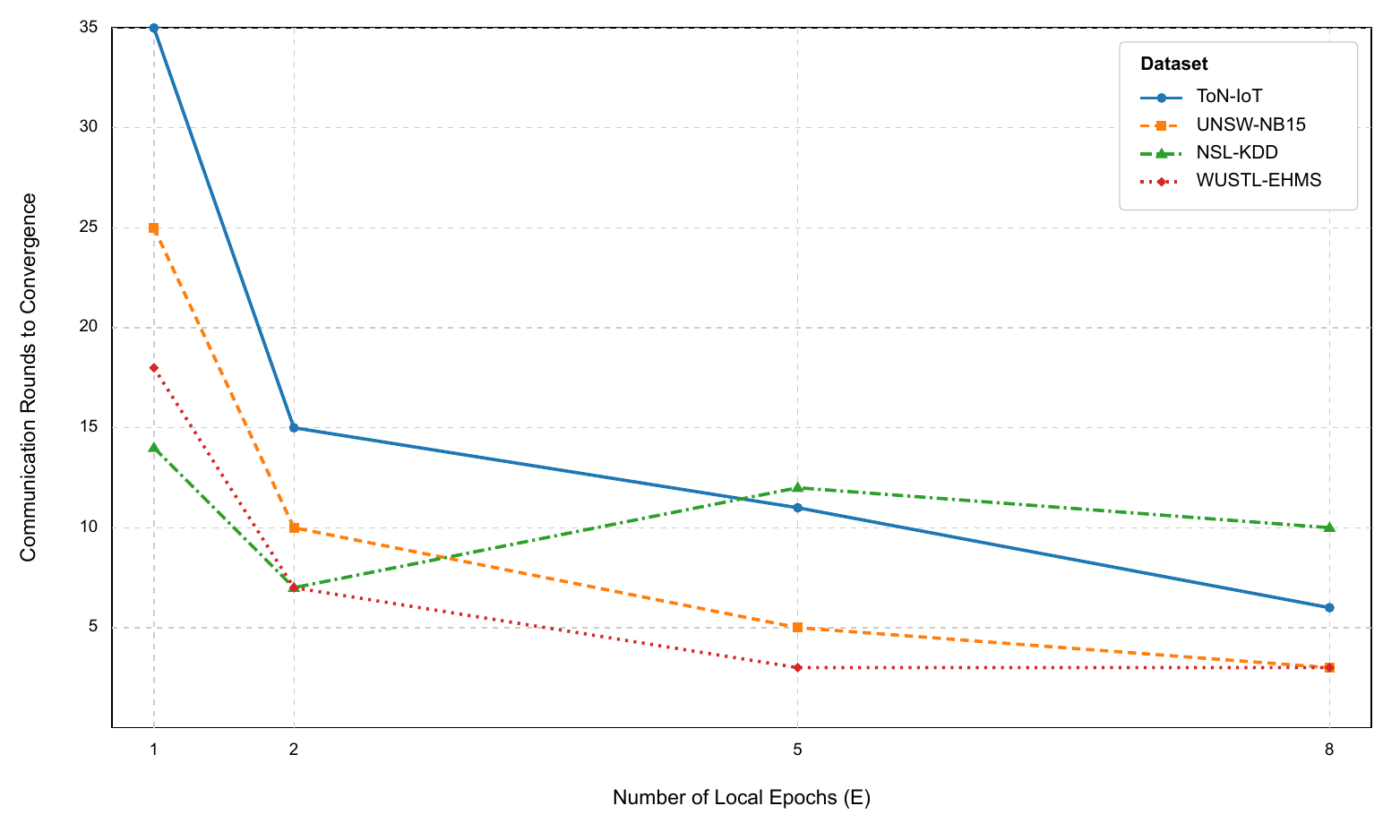}
    \caption{Impact of Local Epochs on Federated Learning Convergence Using Different Datasets}
    \label{Local Epochs}
\end{figure}

\restoregeometry

The influence of these parameters is analyzed with respect to the convergence behavior of the FL process, communication overhead, and intrusion detection performance. This evaluation provides a comprehensive assessment of the capability of the proposed framework to satisfy the operational requirements of practical IoMT deployments while maintaining an appropriate balance between privacy preservation, communication efficiency, and detection effectiveness.

To further enhance the interpretability of the experimental results, the SHAP method is applied to the best-performing models obtained for each dataset. By quantifying the contribution of individual features to each prediction, SHAP provides a transparent interpretation of the feature attribution patterns underlying the decisions of the trained deep learning models. This analysis complements the quantitative evaluation by enabling the identification of the most influential network and medical features involved in anomaly detection, thereby improving the transparency, trustworthiness, and practical interpretability of the proposed intrusion detection framework.

Subsequently, the FL configuration achieving the best overall performance is selected and compared with an equivalent centralized learning approach. This comparison extends beyond conventional predictive performance metrics by jointly evaluating classification performance (Accuracy, Precision, Recall, F1-score, and AUC) and the consistency of SHAP feature attribution patterns between the two learning paradigms. The objective is to assess whether the proposed FL framework preserves not only the detection capability of centralized training but also the same domain-relevant evidence used to support its predictions. Such a comparison provides additional validation that privacy-preserving FL maintains interpretable and reliable feature attribution patterns while preserving the communication-efficient and privacy-preserving properties required for IoMT environments.

\subsubsection{FL Parameters Selections}
The exploration of FL dynamics involves a systematic analysis of key parameters influencing performance and communication cycles. These parameters include the number of participating clients, the fraction fit, and the number of local training epochs.

To evaluate the impact of varying the number of clients on performance and communication cycles, the local epochs are fixed at 1, the fraction fit at 1, and the number of clients is adjusted to 2, 4, 8, and 12. Increasing the number of clients results in a significant reduction in communication cycles required to achieve target performance levels. This trend is evident in the Ton-IoT and NSL-KDD datasets, where the lowest number of communication cycles is achieved with 12 clients, and no significant variation is observed with 8 clients compared to other configurations, as shown in Figure \ref{Client Number} and Tables \ref{NumberClientsToN-IoT} and \ref{NumberClientsNSL-KDD}. This suggests that communication cycles stabilize with 8 or more clients. For the UNSW\_NB15 dataset, the lowest number of communication cycles is achieved with 12 clients, showing a significant reduction compared to 8 clients, a trend supported by Figure \ref{Client Number} and Table \ref{NumberClientsUNSW_NB15}. This aligns with the decreasing trend in communication cycles as the number of clients increases, consistent with observations from the Ton-IoT and NSL-KDD datasets. In contrast, the WUSTL-EHMS dataset exhibits an inverse trend, where increasing the number of clients leads to a higher number of communication rounds to achieve target performance, which is distinctly illustrated in Figure \ref{Client Number} and depicted in Table \ref{NumberClientsWUSTL-EHMS}. This behavior is likely due to the limited sample size in the dataset, resulting in insufficient data distribution across clients and requiring additional communication rounds for model generalization.

\newgeometry{margin=1cm}

\begin{figure}[!htbp]
    \centering

    \begin{subfigure}[t]{0.48\textwidth}
        \centering
        \includegraphics[width=\linewidth]{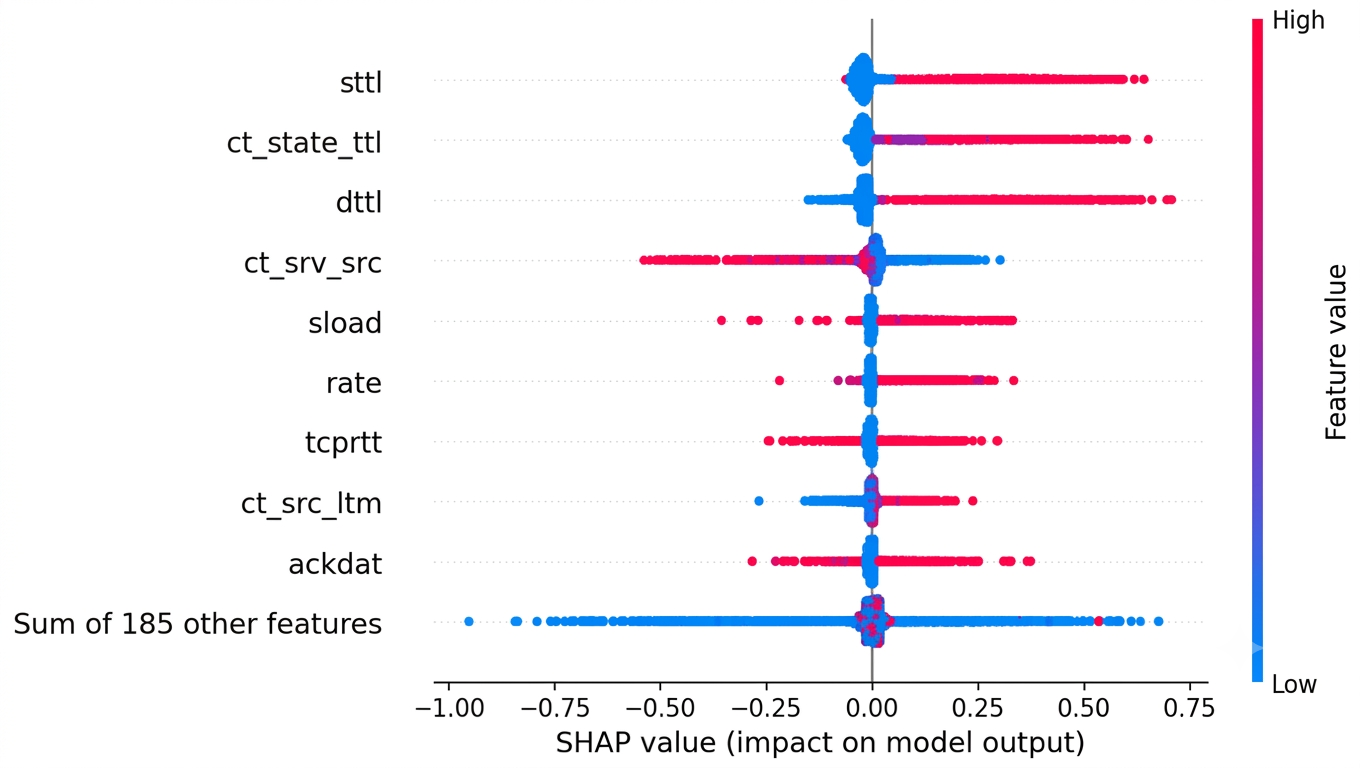}
        \caption{Beeswarm Plot.}
        \label{bsp_UNSW-NB15}
    \end{subfigure}
    \hfill
    \begin{subfigure}[t]{0.48\textwidth}
        \centering
        \includegraphics[
            width=\linewidth,
            keepaspectratio
        ]{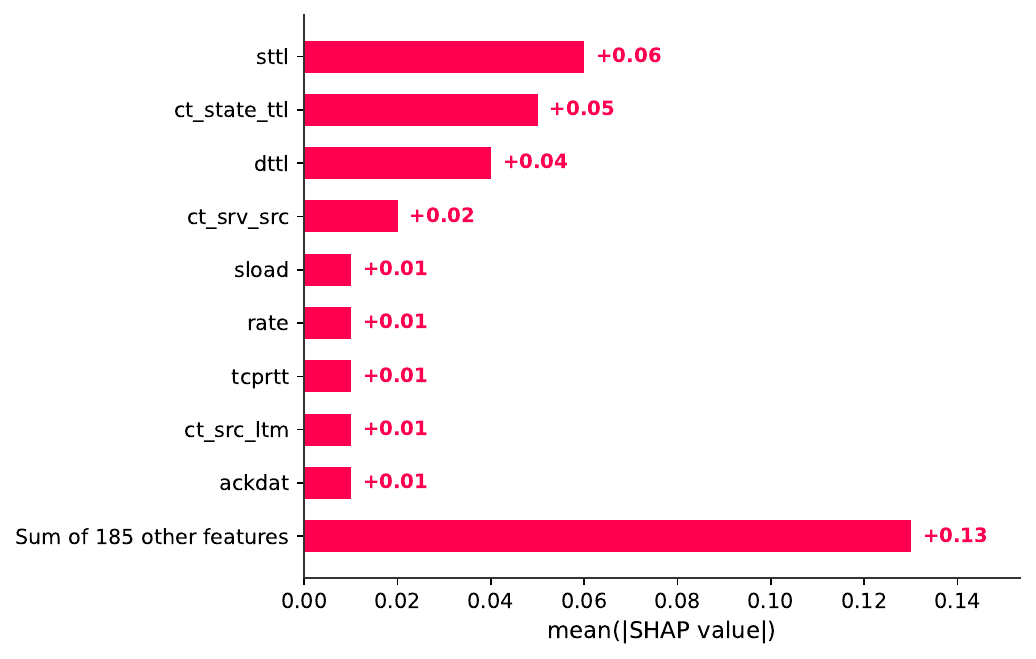}
        \caption{Bar Plot.}
        \label{bpUNSW-NB15}
    \end{subfigure}

    \caption{XAI Results for UNSW-NB15 Dataset Using FL.}
    \label{xaiUNSW-NB15}
\end{figure}

\begin{figure}[!htbp]
    \centering

    \begin{subfigure}[t]{0.48\textwidth}
        \centering
        \includegraphics[width=\linewidth]{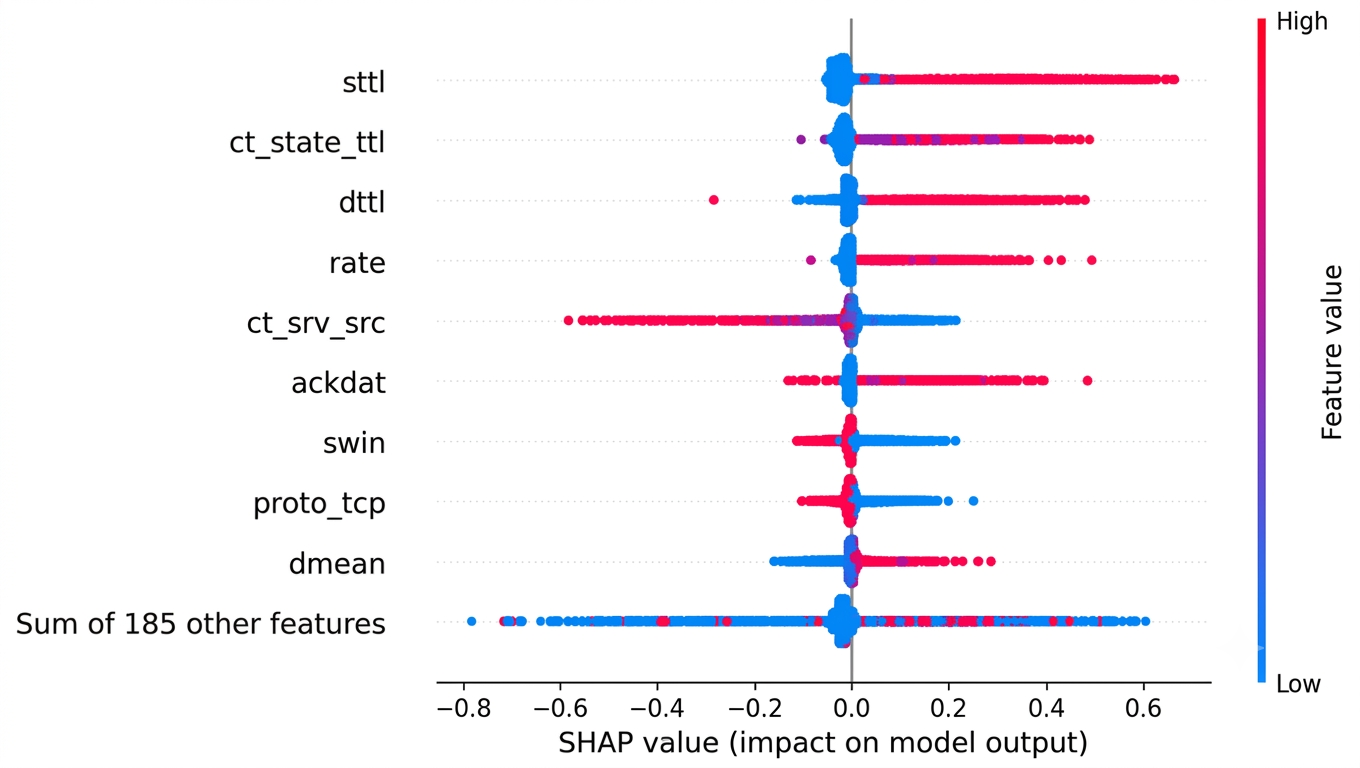}
        \caption{Beeswarm Plot.}
        \label{bsp_UNSW-NB15_centraliser}
    \end{subfigure}
    \hfill
    \begin{subfigure}[t]{0.48\textwidth}
        \centering
        \includegraphics[
            width=\linewidth,
            keepaspectratio
        ]{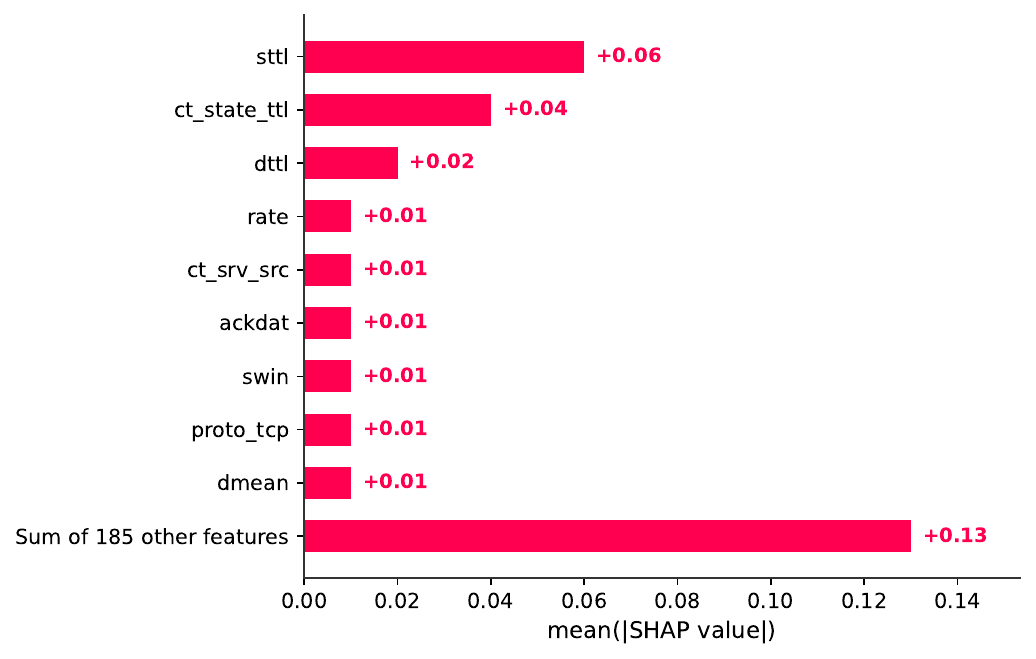}
        \caption{Bar Plot.}
        \label{bp_UNSW-NB15_centraliser}
    \end{subfigure}

    \caption{XAI Results for UNSW-NB15 Dataset Using Centralized Learning.}
    \label{xai_unsw_nb15_centraliser}
\end{figure}

\begin{figure}[!htbp]
    \centering

    \begin{subfigure}[t]{0.48\textwidth}
        \centering
        \includegraphics[width=\linewidth]{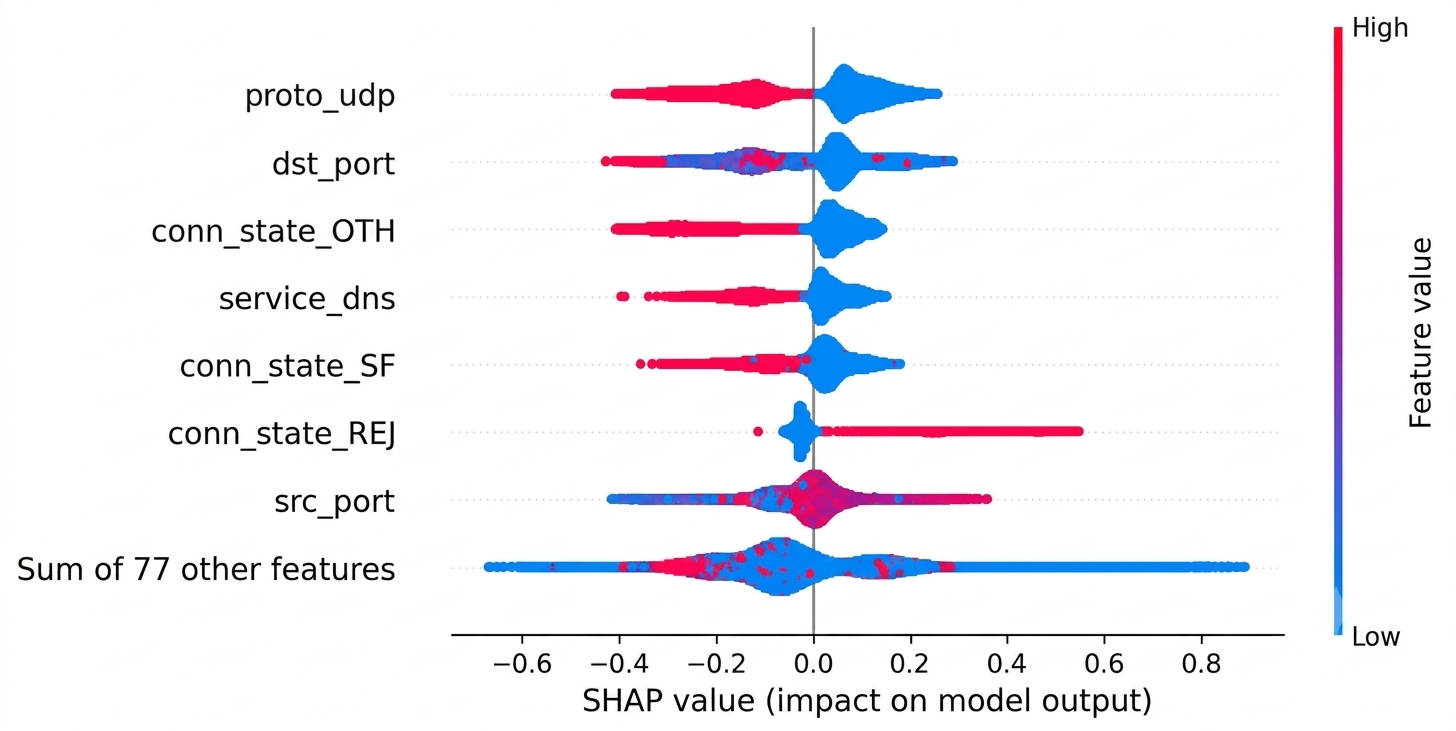}
        \caption{Beeswarm Plot.}
        \label{bsp_ton_iot}
    \end{subfigure}
    \hfill
    \begin{subfigure}[t]{0.48\textwidth}
        \centering
        \includegraphics[
            width=\linewidth,
            keepaspectratio
        ]{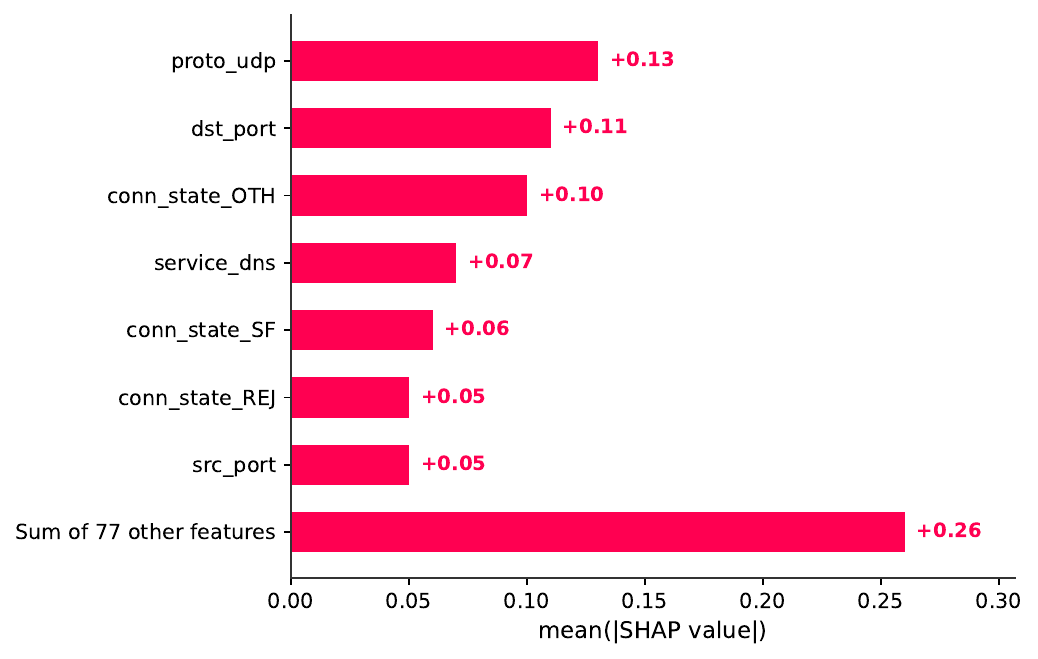}
        \caption{Bar Plot.}
        \label{bp_ton_iot}
    \end{subfigure}

    \caption{XAI Results for ToN-IoT Dataset Using FL.}
    \label{xaiToN-IoT}
\end{figure}

\begin{figure}[!htbp]
    \centering

    \begin{subfigure}[t]{0.48\textwidth}
        \centering
        \includegraphics[width=\linewidth]{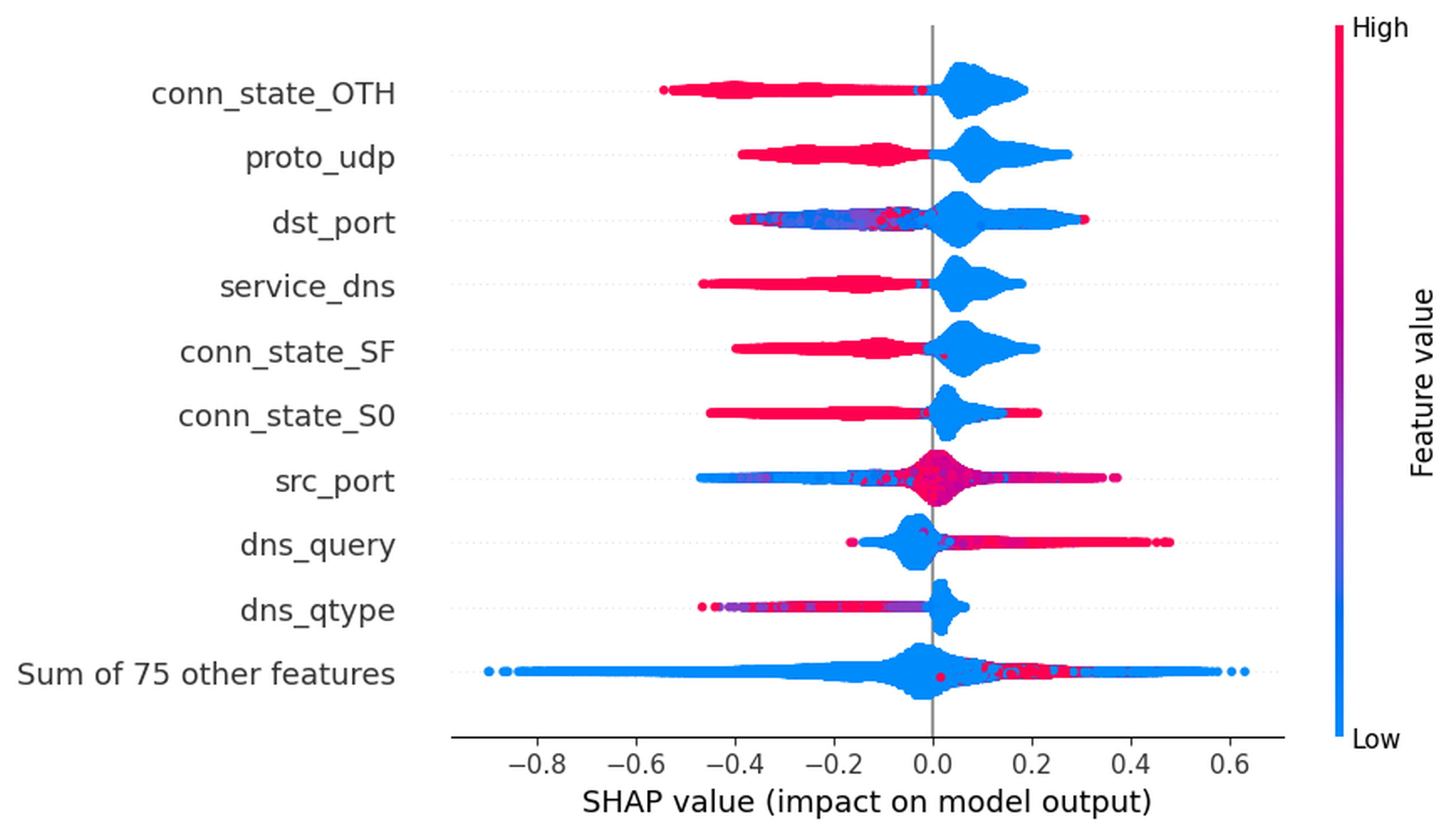}
        \caption{Beeswarm Plot.}
        \label{bsp_ton_iot_centraliser}
    \end{subfigure}
    \hfill
    \begin{subfigure}[t]{0.48\textwidth}
        \centering
        \includegraphics[
            width=\linewidth,
            keepaspectratio
        ]{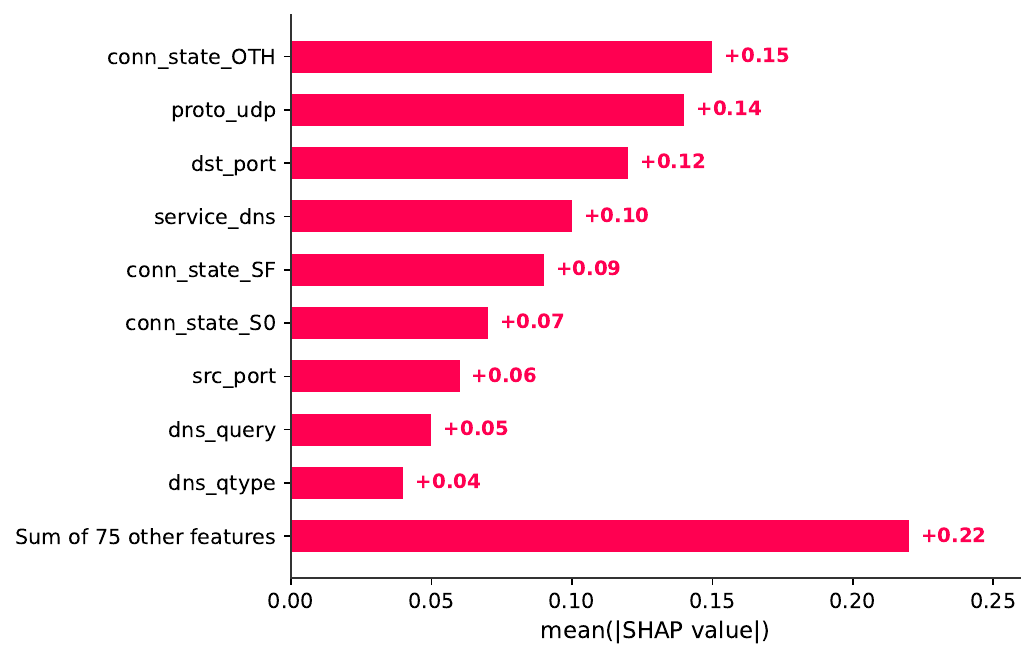}
        \caption{Bar Plot.}
        \label{bp_ton_iot_centraliser}
    \end{subfigure}

    \caption{XAI Results for ToN-IoT Dataset Using Centralized Learning.}
    \label{xai_ton_iot_centraliser}
\end{figure}

\begin{figure}[!htbp]
    \centering

    \begin{subfigure}[t]{0.48\textwidth}
        \centering
        \includegraphics[width=\linewidth]{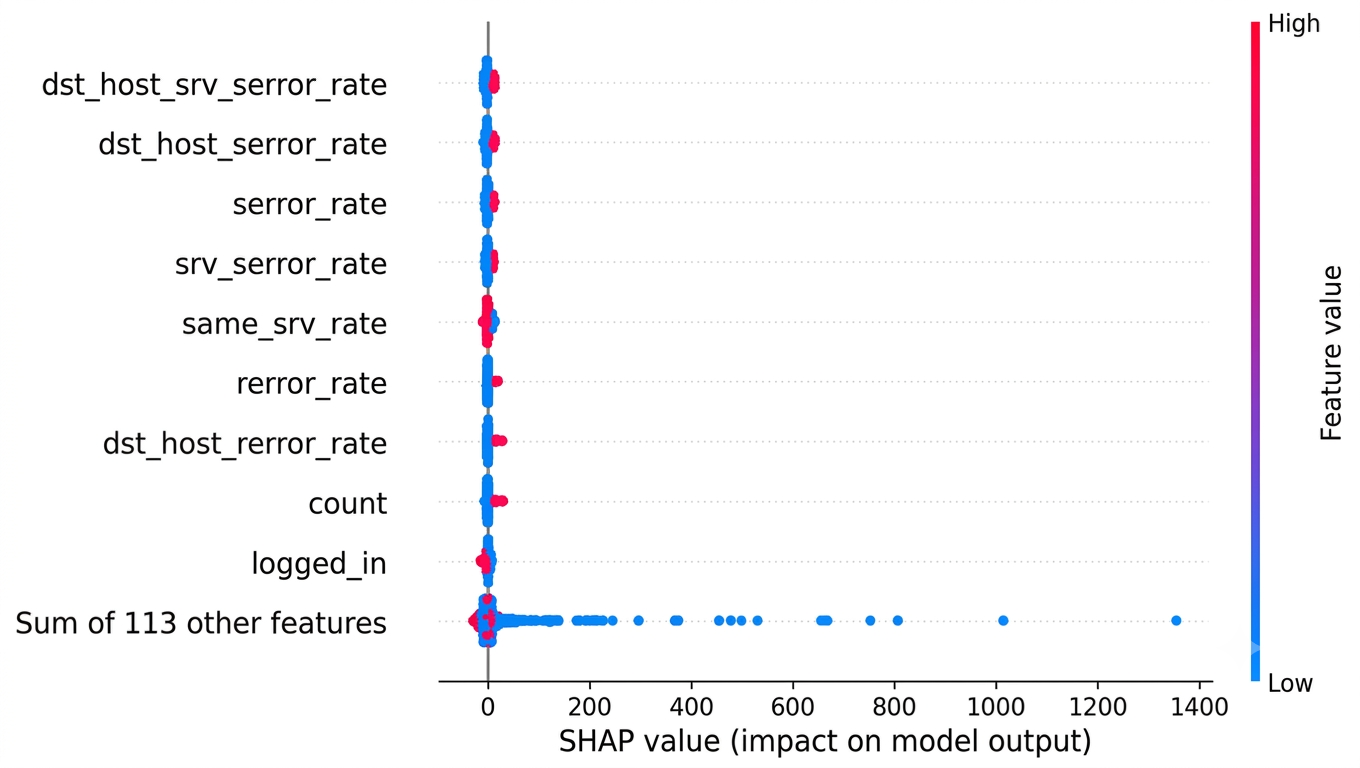}
        \caption{Beeswarm Plot.}
        \label{bsp_NSL-KDD}
    \end{subfigure}
    \hfill
    \begin{subfigure}[t]{0.48\textwidth}
        \centering
        \includegraphics[
            width=\linewidth,
            keepaspectratio
        ]{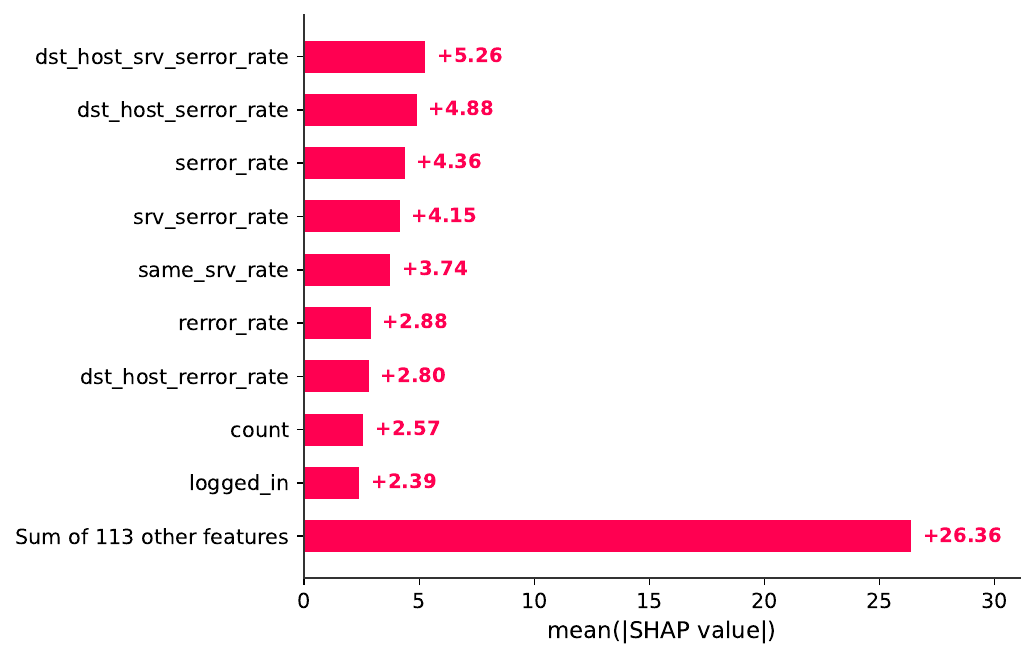}
        \caption{Bar Plot.}
        \label{bp_NSL-KDD}
    \end{subfigure}

    \caption{XAI Results for NSL-KDD Dataset Using FL.}
    \label{xaiNSL-KDD}
\end{figure}

\begin{figure}[!htbp]
    \centering

    \begin{subfigure}[t]{0.48\textwidth}
        \centering
        \includegraphics[width=\linewidth]{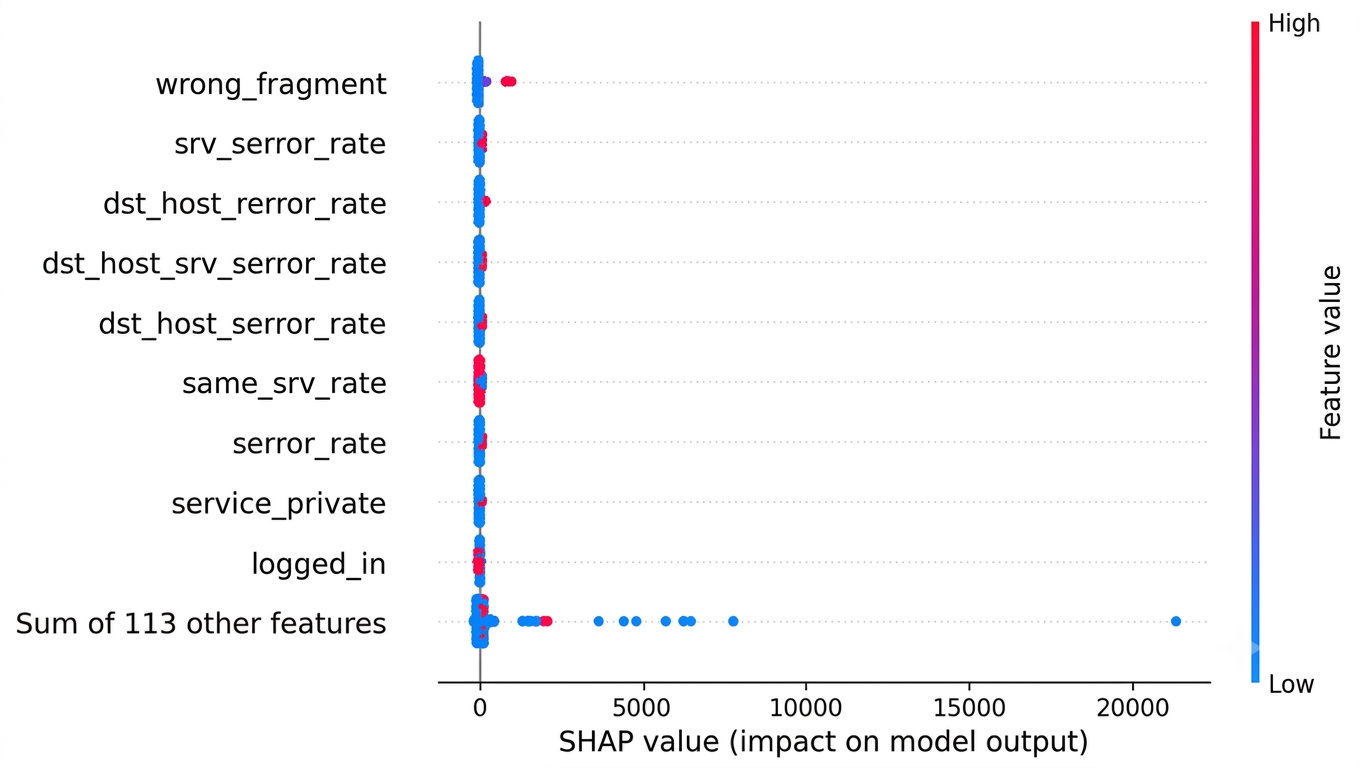}
        \caption{Beeswarm Plot.}
        \label{bsp_NSL-KDD_centraliser}
    \end{subfigure}
    \hfill
    \begin{subfigure}[t]{0.48\textwidth}
        \centering
        \includegraphics[
            width=\linewidth,
            keepaspectratio
        ]{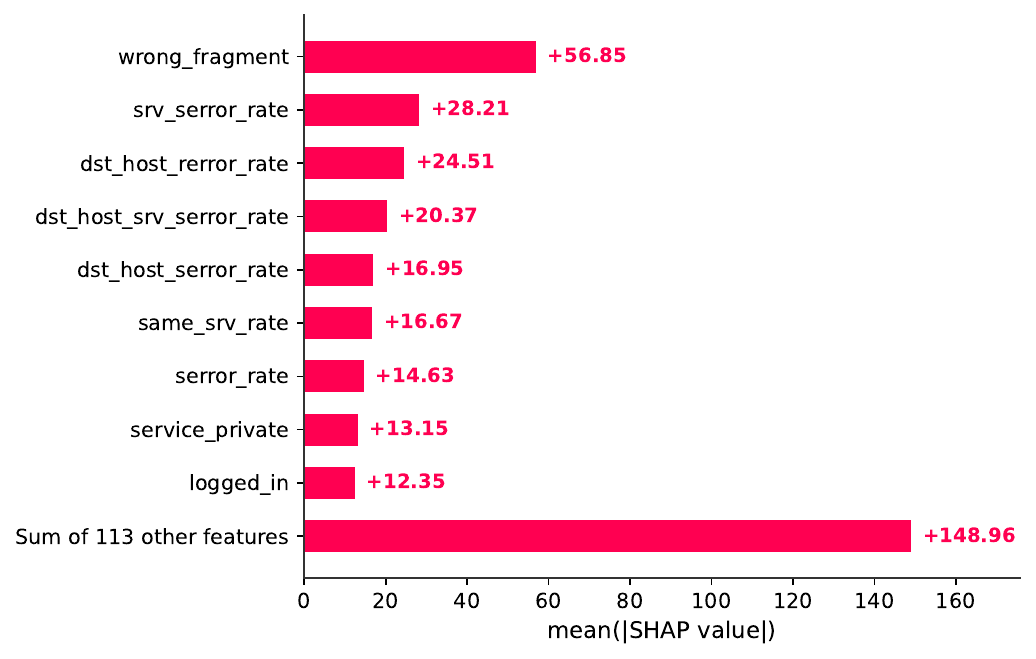}
        \caption{Bar Plot.}
        \label{bp_NSL-KDD_centraliser}
    \end{subfigure}

    \caption{XAI Results for NSL-KDD Dataset Using Centralized Learning.}
    \label{xai_nsl_kdd_centraliser}
\end{figure}

\begin{figure}[!htbp]
    \centering

    \begin{subfigure}[t]{0.48\textwidth}
        \centering
        \includegraphics[width=0.75\linewidth]{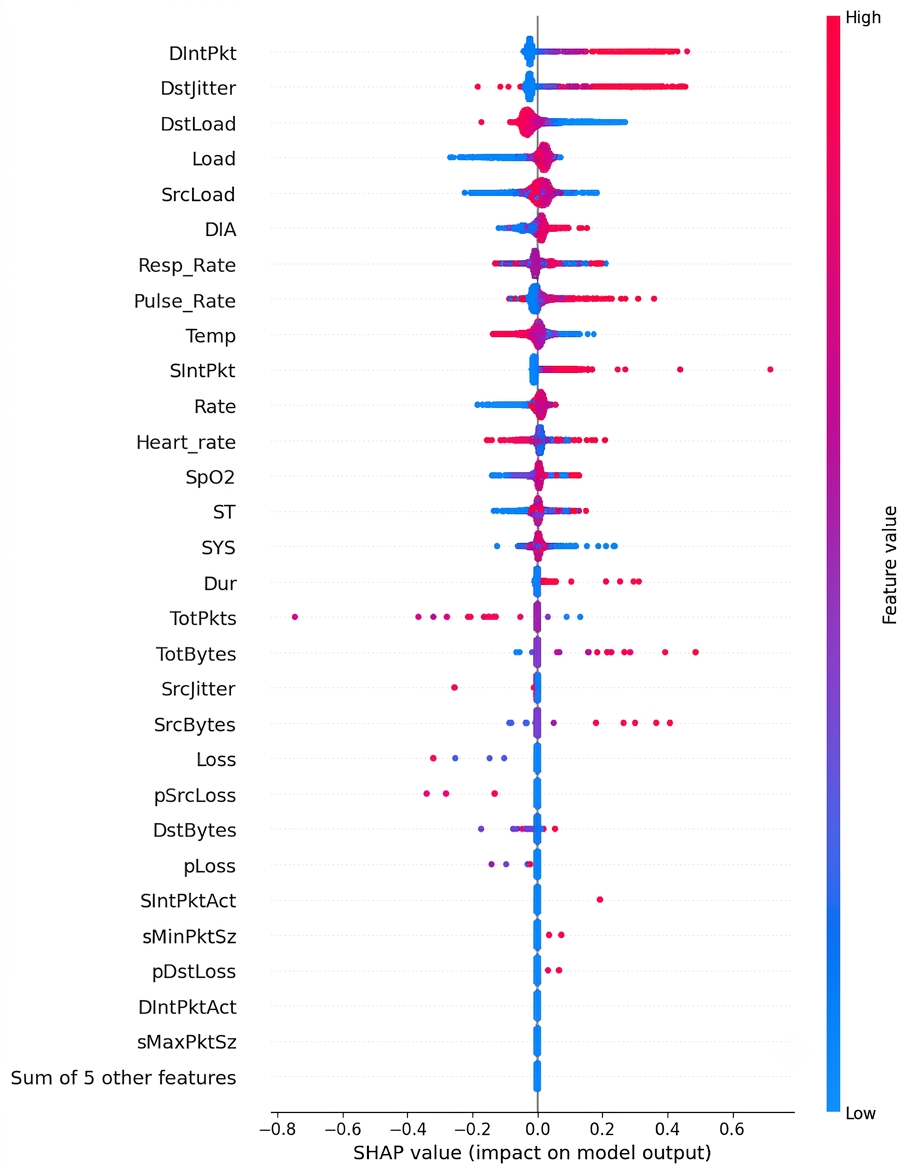}
        \caption{Beeswarm Plot.}
        \label{bsp_wustl}
    \end{subfigure}
    \hfill
    \begin{subfigure}[t]{0.48\textwidth}
        \centering
        \includegraphics[
            width=\linewidth,
            keepaspectratio
        ]{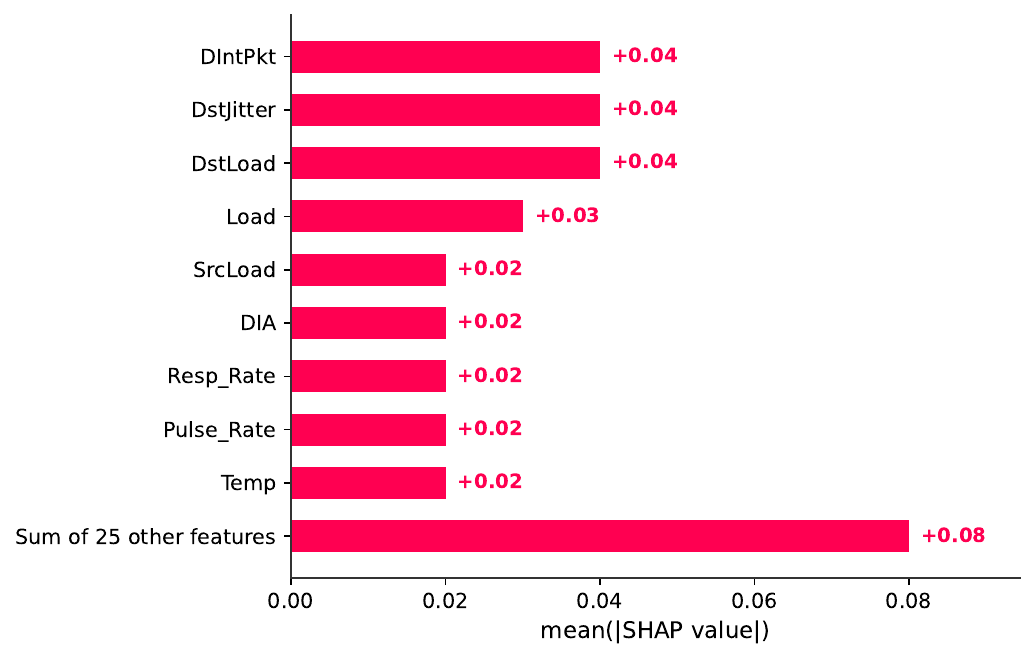}
        \caption{Bar Plot.}
        \label{bp_wustl}
    \end{subfigure}

    \caption{XAI Results for WUSTL-EHMS Dataset Using FL.}
    \label{xaiWUSTL-EHMS}
\end{figure}

\begin{figure}[!htbp]
    \centering

    \begin{subfigure}[t]{0.48\textwidth}
        \centering
        \includegraphics[width=0.75\linewidth]{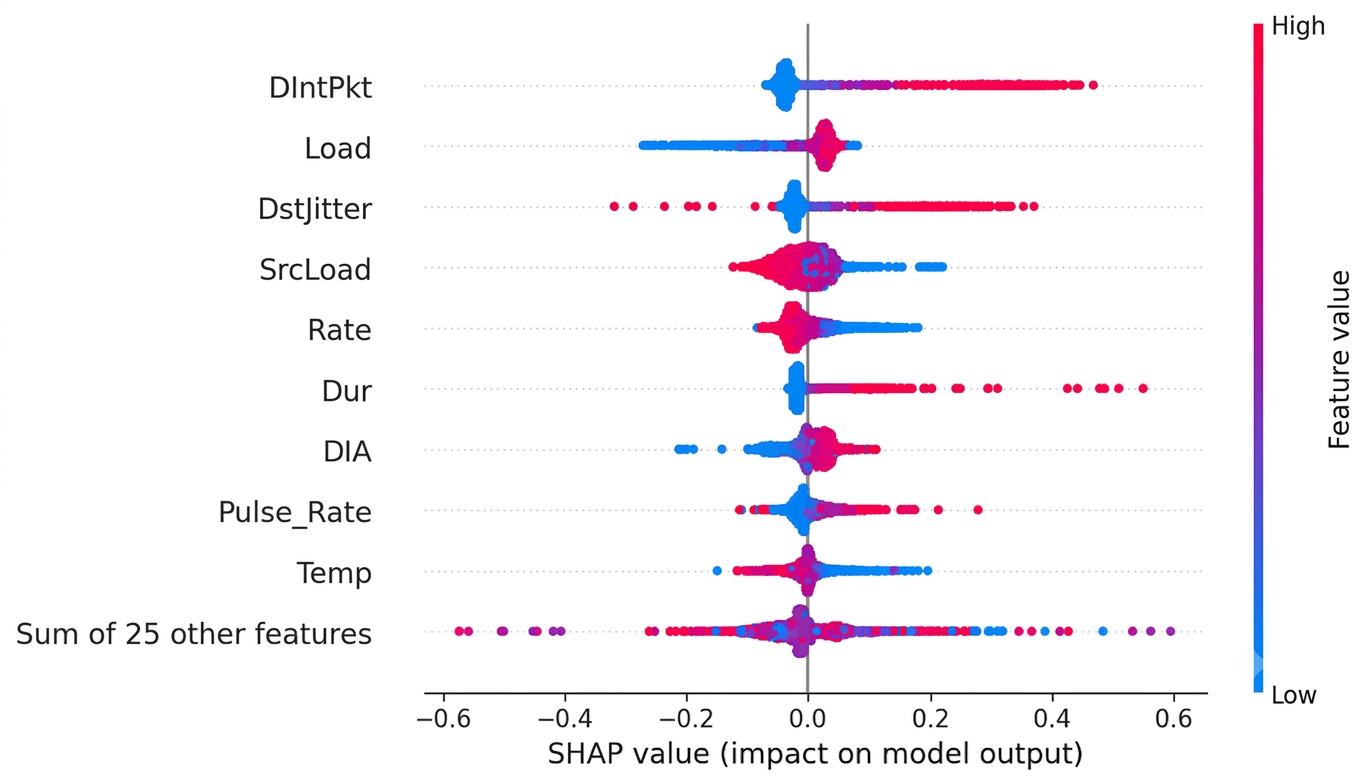}
        \caption{Beeswarm Plot.}
        \label{bsp_wustl_centraliser}
    \end{subfigure}
    \hfill
    \begin{subfigure}[t]{0.48\textwidth}
        \centering
        \includegraphics[
            width=\linewidth,
            keepaspectratio
        ]{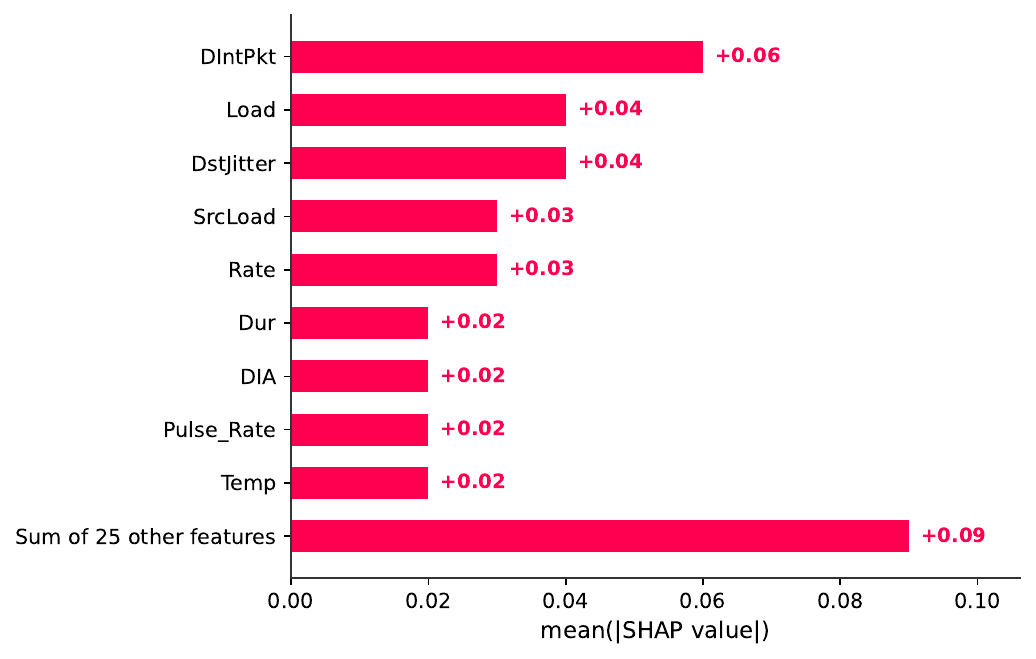}
        \caption{Bar Plot.}
        \label{bp_wustl_centraliser}
    \end{subfigure}

    \caption{XAI Results for WUSTL-EHMS Dataset Using Centralized Learning.}
    \label{xai_wustl_ehms_centraliser}
\end{figure}

\restoregeometry

Next, the analysis examines the influence of the fraction fit parameter on performance and communication cycles. For this evaluation, the number of local epochs is set to 1, and the number of clients is fixed at 8 for the Ton-IoT and NSL-KDD datasets, as communication rounds decrease and stabilize at this client count. Similarly, the WUSTL-EHMS dataset is tested with 8 clients, despite this not being the optimal configuration, to better simulate real-world FL scenarios. For the UNSW-NB15 dataset, the number of clients is fixed at 12, as this configuration minimizes communication rounds. The fraction fit is then varied to 0.1, 0.5, and 1 to assess its impact.

Increasing the fraction fit effectively reduces communication cycles while maintaining target performance levels. This trend is observed in the Ton-IoT and NSL-KDD datasets, as illustrated in Figure \ref{Fraction} and Tables \ref{FractionFitton_iot} and \ref{FractionFitNSL-KDD}. However, for the UNSW-NB15 and WUSTL-EHMS datasets, the lowest number of communication rounds is achieved with a fraction fit of 0.5, with a slight increase observed at a fraction fit of 1, a non-monotonic relationship clearly shown in Figure \ref{Fraction} and detailed in Tables \ref{FractionFitUNSW_NB15} and \ref{FractionFitWUSTL-EHMS}. This confirms the general trend of reduced communication rounds with higher fraction fit values, though the optimal value can be dataset-dependent.

Finally, the analysis investigates the impact of the number of local epochs on performance and communication cycles. The fraction fit is set to 1, as this configuration is shown to minimize communication rounds. The number of local epochs is varied to 1, 2, 5, and 8.

A notable finding is the consistent reduction in communication cycles with an increase in local epochs while maintaining strong performance. 
This trend is observed across most datasets, as clearly depicted in Figure \ref{Local Epochs} and shown in Tables \ref{LocalEpochston_iot}, \ref{LocalEpochsUNSW_NB15}, and \ref{LocalEpochsWUSTL-EHMS}. However, the NSL-KDD dataset achieves the optimal number of communication rounds with only 2
epochs, a deviation from the overall trend that is visually apparent in Figure \ref{Local Epochs} and illustrated in Table \ref{LocalEpochsNSL-KDD}, though the results remain close to those obtained with 8 epochs. This further supports the overall trend of reduced communication rounds with increased local epochs, while also highlighting dataset-specific idiosyncrasies.

\subsubsection{XAI results}

The majority of research conducted in the field of IDS-based ML focuses on performance aspects, neglecting the explanatory side of ML/DL models, which lack transparency and trust. XAI provides a solution to explain complex models, enabling the identification of issues and validating the accuracy of ML models for threat detection. This helps administrators and security analysts gain a better understanding of the model's reasoning.

Previous test results demonstrate the IDS system's commendable performance in anomaly detection. However, an investigation into why the proposed solution predicts as it does is imperative. To address this concern, the SHAP method is employed, enabling the identification of feature relevance in anomaly detection. The outcomes of the SHAP method can be depicted graphically, utilizing variable-length bars and color coding called beeswarm plots. This visualization effectively demonstrates how each level or range of values of a particular feature positively or negatively influences the classification result.

Each point on the graph represents a feature value. Red points denote higher feature values, while blue points indicate lower feature values. Values on the left side of the x-axis tend toward the normal class, whereas those on the right side tend toward the anomaly class \cite{gurbuz2023explainable}. There are also other types of graphs, such as bar plots. In these plots, the x-axis represents the Shapley value, and the y-axis represents the feature names. Features with the most significant impact are positioned at the top of the graph, while those with the least impact are at the bottom \cite{aljuhani2023intelligent}.

In this approach, the SHAP value is applied to the final FL model, achieving the best performance using the test dataset. This process produces two types of graphics: beeswarm plots and bar plots.

\newgeometry{margin=1cm}

\begin{sidewaysfigure}[!p]
    \centering

    \begin{subfigure}[t]{0.48\textheight}
        \centering
        \includegraphics[
            width=\textwidth,
            height=0.80\textheight,
            keepaspectratio,
            trim=0.5cm 0.5cm 0.5cm 0.5cm,
            clip
        ]{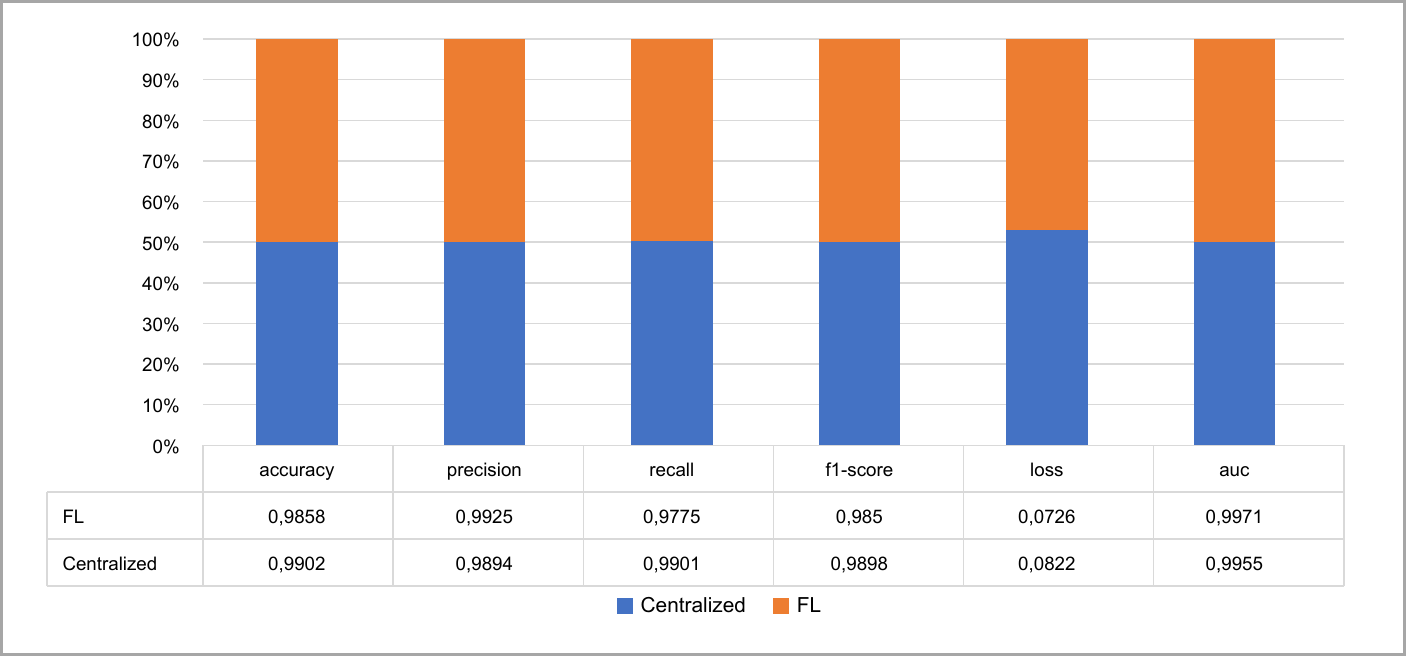}
        \caption{NSL-KDD dataset}
        \label{NSL-KDD cent vs fl}
    \end{subfigure}
    \hfill
    \begin{subfigure}[t]{0.48\textheight}
        \centering
        \includegraphics[
            width=\textwidth,
            height=0.80\textheight,
            keepaspectratio,
            trim=0.5cm 0.5cm 0.5cm 0.5cm,
            clip
        ]{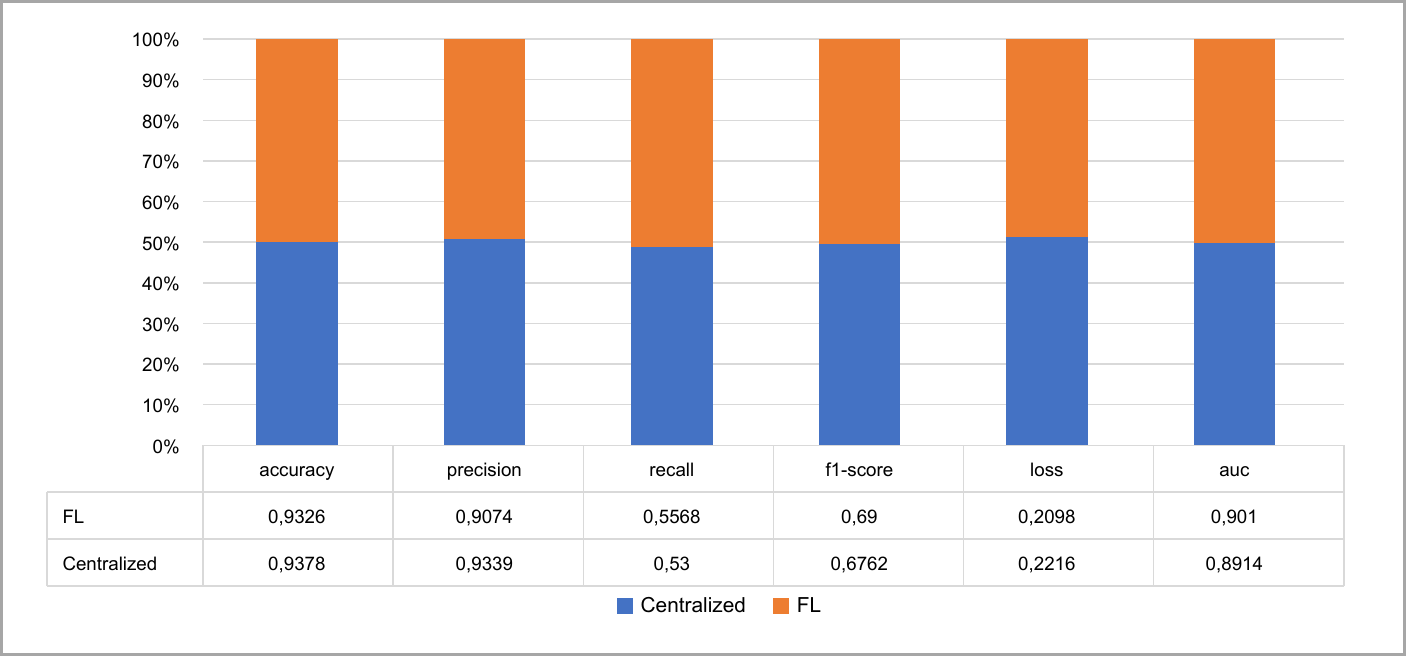}
        \caption{WUSTL-EHMS dataset}
        \label{WUSTL-EHMS cent vs fl}
    \end{subfigure}

    \vspace{0.7cm}

    \begin{subfigure}[t]{0.48\textheight}
        \centering
        \includegraphics[
            width=\textwidth,
            height=0.80\textheight,
            keepaspectratio,
            trim=0.5cm 0.5cm 0.5cm 0.5cm,
            clip
        ]{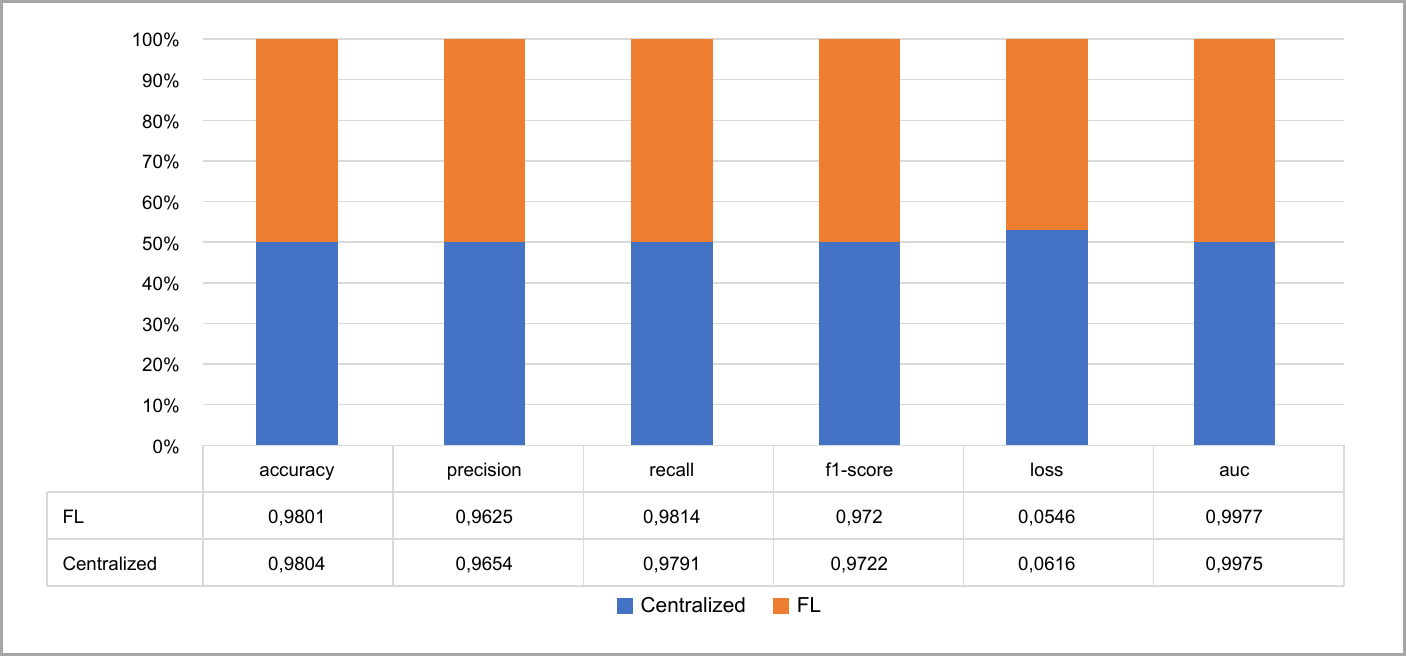}
        \caption{TON\_IOT dataset}
        \label{TON_IOT cent vs fl}
    \end{subfigure}
    \hfill
    \begin{subfigure}[t]{0.48\textheight}
        \centering
        \includegraphics[
            width=\textwidth,
            height=0.80\textheight,
            keepaspectratio,
            trim=0.5cm 0.5cm 0.5cm 0.5cm,
            clip
        ]{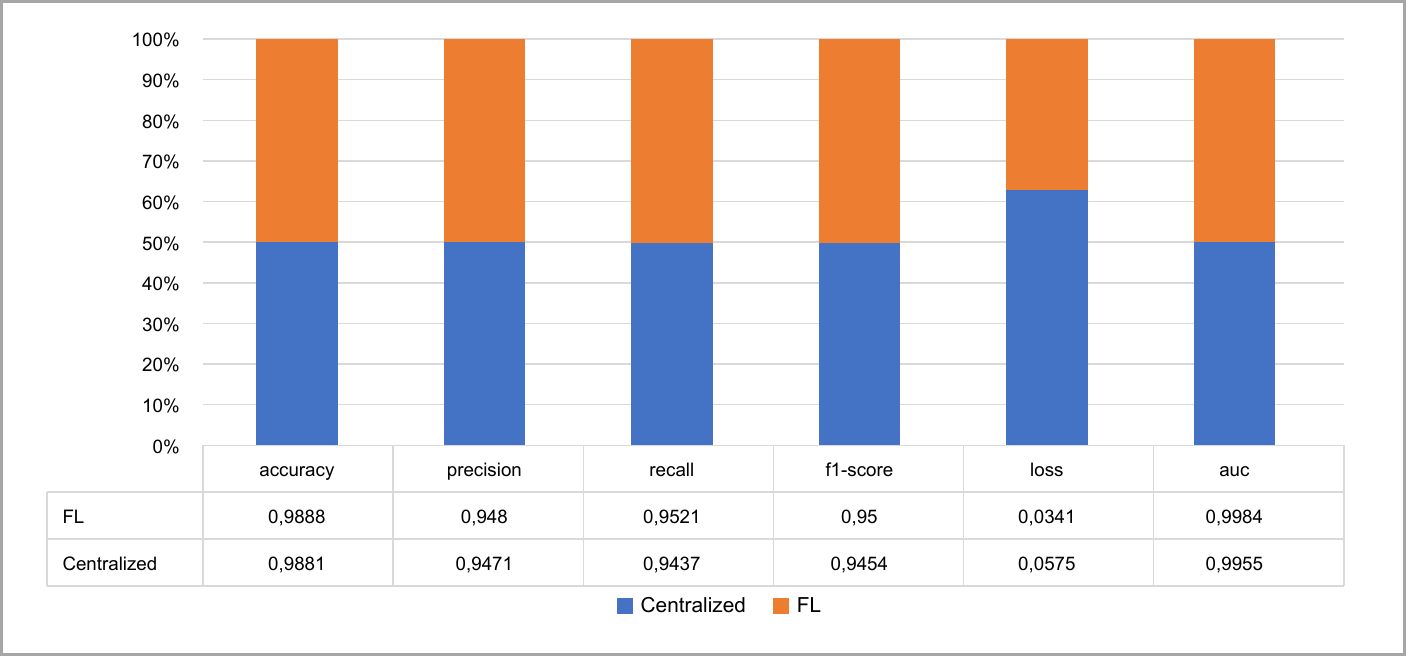}
        \caption{UNSW\_NB15 dataset}
        \label{UNSW_NB15 cent vs fl}
    \end{subfigure}

    \caption{Centralized and FL Performances}
    \label{Centralized and FL Performances}

\end{sidewaysfigure}

\restoregeometry

The SHAP results for the UNSW-NB15 dataset are depicted in Figure \ref{xaiUNSW-NB15}, while the key features exerting the most significant impact on the anomaly detection classification process, as shown in Figure \ref{bpUNSW-NB15}, include sttl (Time to Live from the source to the destination), ct\_state\_ttl (Connection state value of Time to Live), and dttl (Time to Live from the destination to the source). Notably, high values of these features, all linked to Time To Live (TTL), play a pivotal role in anomaly detection, visually represented in the accompanying Figure \ref{bsp_UNSW-NB15}.

Prominent patterns emerge, underscoring that elevated values in sttl, ct\_state\_ttl, and dttl serve as indicators of anomalies, especially concerning TTL. Under normal circumstances, TTL values tend to exhibit stability or fall within specific ranges. However, higher values may signify abnormal extensions of connections, potentially employed to circumvent temporary security mechanisms. It is essential to note that certain attacks comprise very few samples, posing a challenge for ML in discerning crucial features for the detection of such anomalies.

In the ToN-IoT dataset, as illustrated in Figure \ref{xaiToN-IoT}, the features wielding the most significant impact on anomaly detection classification include proto \_udp (indicating the use of UDP), dst\_port (representing destination ports), conn\_state\_oth (denoting other unspecified connection states), service\_dns (reflecting the use of DNS services), conn\_state\_SF (indicating an established connection with successful data exchange), conn\_state\_rej (signifying connection rejection or inability to establish), and src\_port (depicting source ports from the endpoint's TCP/UDP ports), as shown in Figure \ref{bp_ton_iot}.
The absence of UDP protocol usage, as demonstrated in Figure \ref{bsp_ton_iot}, is pivotal in detecting anomalies, emphasizing its role in discerning abnormal network behavior. Notably, lower destination port numbers indicate ongoing attacks, with attackers often targeting well-known services associated with such ports (e.g., FTP=21, SSH=22, or HTTP=80).
Correlated attributes, namely conn\_state\_oth, conn\_state\_SF, and conn\_state\_rej, play a critical role in attack detection. The absence of a specific connection state in conn\_state\_oth, as depicted in Figure \ref{bsp_ton_iot}, suggests potential interception or falsification of traffic, while conn\_state\_SF points to established connections without successful data exchange, possibly indicating suspicious activities. Simultaneously, conn\_state\_rej highlights repeated or massive attempts to establish rejected connections, hinting at DDoS attacks or vulnerability exploitation.
The absence of DNS port usage, as shown in Figure \ref{bsp_ton_iot}, indicates a potential threat, as attackers may manipulate DNS requests to mask their activities. Additionally, elevated values of the source port attribute enhance attack detection, as attackers often initiate attacks from dynamic ports not assigned to well-known services like FTP.

For the NSL-KDD dataset, SHAP results are illustrated in Figure \ref{xaiNSL-KDD}. The crucial features dictating the anomaly detection classification process include dst\_host\_srv\_serror\_rate, dst\_host\_serror\_rate, serror\_rate, and srv\_serror\_rate  as demonstrated in Figure \ref{bp_NSL-KDD}. These features bear unique significance, delineating various aspects of connection behavior. Notably, dst\_host\_srv\_serror\_rate quantifies the percentage of connections activating specific flags among those aggregated in dst\_host\_srv\_count, while dst\_host\_serror\_rate gauges the same metric within dst\_host\_count. Similarly, serror\_rate represents the percentage of connections with specific flags activated among those aggregated in count, and srv\_serror\_rate measures the percentage of connections featuring SYN errors.
A comprehensive analysis of Figure \ref{bsp_NSL-KDD} reveals a compelling trend where elevated values of dst\_host\_srv\_serror\_rate, dst\_host\_serror\_rate, serror\_rate, and srv\_serror\_rate significantly contribute to the detection of attacks. These features act as key indicators of anomalies, signaling a substantial increase in errors or failures within connections. This heightened activity may indicate attacks aiming to overwhelm target system resources, potentially rendering them unavailable. Furthermore, these anomalies could signify unauthorized access attempts, exploitation of known vulnerabilities, or even reconnaissance and probing of the network.
It's noteworthy that while these features prove highly relevant for various attack types, they exhibit comparatively less relevance for U2R-type attacks. This observation can be attributed to the limited number of samples available for such attacks.

The SHAP results for the WUSTL-EHMS dataset, as illustrated in Figure \ref{xaiWUSTL-EHMS}, highlight features influencing the classification process in anomaly detection. These features, outlined in Figure \ref{bp_wustl}, include Dintpkt (Destination Inter Packet), dstjitter (Destination Jitter), dstload (Destination Load), load, and srcload (Source Load). Each of these features holds distinct relevance for understanding and characterizing network behavior, especially in the context of health-related host data.
Significantly, features associated with health data have a profound impact on detecting attacks, particularly those related to data injection. Anomalies in health-related metrics can be identified by values deviating from typical ranges. For example, elevated pulse\_rate or unusually low temperatures, as depicted in Figures \ref{bsp_wustl}, can signal abnormal conditions.
Elevated values of Dintpkt or dstjitter, which are correlated features, may indicate unusual intervals or abnormal temporal variations between packets or data received by the destination. In the context of spoofing, such variations could signal an attempt to manipulate network traffic, concealing or altering the true origin of packets. This observation aligns with the understanding that packets in spoofing scenarios are often manually modified by attackers, sent individually with extended time intervals between them, rather than in an organized flow.

Low values of dstload, load, and srcload also play a significant role in anomaly detection. A destination with  low load may be unusual in a normal context, potentially indicating an attacker not targeting system availability but engaging in other types of attacks, such as injection or scanning attacks. This behavior might be explained by the attacker's intent to remain inconspicuous and avoid detection, highlighting the multifaceted role these features play in discerning anomalies within network behavior.

\subsubsection{Centralized vs FL :}

In the context of comparing the centralized and federated approaches, parameters are meticulously selected to achieve optimal performance for FL, and these outcomes are compared with those of the centralized approach. Comparative tests are conducted on diverse datasets, relying on metrics such as accuracy, precision, recall, F1 score, and AUC.

The results are consolidated into stacked bar plots, as depicted in Figure \ref{Centralized and FL Performances}, where FL results are represented in orange, and centralized results are represented in blue. The outcomes on the NSL-KDD, TON-IOT, UNSW\_NB15, and WUSTL-EHMS datasets, presented in Tables \ref{NSL-KDD cent vs fl}, \ref{TON_IOT cent vs fl}, \ref{UNSW_NB15 cent vs fl}, and \ref{WUSTL-EHMS cent vs fl} respectively, show bars being halved, indicating comparable results between FL and centralized approaches. Consequently, training models on separate data partitions generalizes as effectively as training a global model on the entire dataset. This dual advantage of achieving comparable detection outcomes while preserving data privacy highlights the efficacy and privacy-centric nature of the FL-based IDS, positioning it as a robust and privacy-aware solution for intrusion detection.

In addition to the comparable predictive performance, the SHAP analysis provides complementary evidence supporting the consistency of the decision-making process under both centralized and FL paradigms. Figures~\ref{xaiUNSW-NB15}--\ref{xai_wustl_ehms_centraliser} demonstrate that both models consistently identify the same domain-relevant features as the primary contributors to attack detection across all evaluated datasets. Figures~\ref{bpUNSW-NB15} and \ref{bp_UNSW-NB15_centraliser} show that, for the UNSW-NB15 dataset, both learning paradigms consistently assign the highest importance to TTL-related attributes, namely \textit{sttl}, \textit{ct\_state\_ttl}, and \textit{dttl}. Likewise, Figures~\ref{bp_ton_iot} and \ref{bp_ton_iot_centraliser} demonstrate that, for the ToN-IoT dataset, both models predominantly rely on protocol- and connection-state-related features, including \textit{proto\_udp}, \textit{dst\_port}, \textit{service\_dns}, \textit{conn\_state\_OTH}, \textit{conn\_state\_SF}, and \textit{src\_port}. Similarly, Figures~\ref{bp_NSL-KDD} and \ref{bp_NSL-KDD_centraliser} indicate that, for the NSL-KDD dataset, both learning paradigms assign the greatest importance to connection error-related attributes, including \textit{dst\_host\_srv\_serror\_rate}, \textit{dst\_host\_serror\_rate}, \textit{serror\_rate}, and \textit{srv\_serror\_rate}. Finally, Figures~\ref{bp_wustl} and \ref{bp_wustl_centraliser} reveal that, for the WUSTL-EHMS dataset, both models consistently emphasize traffic- and health-related attributes, including \textit{DIntPkt}, \textit{DstJitter}, \textit{Load}, \textit{SrcLoad}, and \textit{Pulse\_Rate}.

Figures~\ref{bsp_UNSW-NB15} and \ref{bsp_UNSW-NB15_centraliser}, Figures~\ref{bsp_ton_iot} and \ref{bsp_ton_iot_centraliser}, Figures~\ref{bsp_NSL-KDD} and \ref{bsp_NSL-KDD_centraliser}, and Figures~\ref{bsp_wustl} and \ref{bsp_wustl_centraliser} further show that these common features exhibit highly consistent attribution patterns under both learning paradigms. Features that positively contribute to attack prediction in the centralized model retain the same positive influence in the federated model, whereas features associated with normal traffic preserve their negative contribution. The observed differences are therefore limited to minor variations in feature ranking and attribution magnitude, rather than changes in the set of discriminative features or in the direction of their contributions.

Overall, these findings demonstrate that FL preserves the feature attribution patterns observed under centralized learning, indicating that the decentralized training process does not alter the underlying decision rationale of the intrusion detection model. The strong agreement between the SHAP explanations obtained from both learning paradigms across all evaluated datasets further confirms that the proposed federated framework maintains the interpretability of the centralized model while simultaneously providing the privacy and security advantages of distributed learning.

\section{Comparison with Previous Work} \label{Comparison with Previous Work}

The proposed approach advances existing intrusion detection solutions for IoMT environments by combining privacy-preserving FL, XAI, and a comprehensive evaluation conducted under operational conditions representative of distributed healthcare infrastructures. A comparative summary of these characteristics is presented in Table \ref{tab:Comparison_prev_works}.

Unlike many existing studies that rely on a single benchmark dataset or focus exclusively on either network traffic or medical data, the proposed framework is validated using multiple datasets encompassing both medical and network traffic. This heterogeneous evaluation enables the framework to be assessed against a broad range of cyberattacks targeting the confidentiality, integrity, and availability of IoMT systems, thereby providing a more comprehensive validation of its applicability across diverse healthcare deployment scenarios.

From a FL perspective, most existing FL-based intrusion detection studies primarily exploit FL as a privacy-preserving training paradigm, while providing limited discussion of how the FL process should be configured to accommodate the operational characteristics of IoMT environments. In contrast, this work investigates the influence of three key FL parameters—the number of participating clients, the client participation fraction, and the number of local training epochs—by relating each of them to practical IoMT deployment constraints. Specifically, the number of participating clients reflects the scalability of collaborative learning as the population of connected medical devices increases. The client participation fraction represents the dynamic availability of personal devices caused by mobility, intermittent wireless connectivity, battery limitations, and resource constraints, allowing the learning process to reflect realistic participation conditions. The number of local training epochs characterizes the trade-off between communication frequency and local model refinement, directly affecting network utilization in communication-constrained IoMT infrastructures. By analyzing the influence of these parameters on convergence behavior, communication overhead, and intrusion detection performance, the proposed study provides practical insights into the deployment of FL in IoMT systems beyond the performance-oriented evaluations commonly reported in previous FL-based intrusion detection approaches.

Regarding model interpretability, many existing intrusion detection approaches either do not incorporate XAI mechanisms or rely on LIME, which primarily provides local explanations for individual predictions. In contrast, the proposed framework employs SHAP to provide a comprehensive global interpretation of the trained intrusion detection model while also enabling the analysis of individual predictions when needed. The generated explanations are further enhanced through visualization techniques, including beeswarm and bar plots, together with a detection history interface that facilitates the interpretation of security events by both technical and non-technical stakeholders.

Finally, the proposed framework explicitly addresses the ethical and regulatory requirements associated with healthcare cybersecurity. By combining FL with XAI, it preserves the confidentiality of patient data while improving the transparency and auditability of intrusion detection decisions. These characteristics support compliance with healthcare data protection regulations, including GDPR and HIPAA, while promoting fairness, accountability, and trust in AI-assisted security monitoring for distributed IoMT infrastructures.

\begin{table}
    \begin{tabular}{|>{\centering\arraybackslash}p{2cm}|
                    >{\centering\arraybackslash}p{2cm}|
                    >{\centering\arraybackslash}p{2cm}|
                    >{\centering\arraybackslash}p{2cm}|
                    >{\centering\arraybackslash}p{2cm}|
                    >{\centering\arraybackslash}p{2cm}|
                    >{\centering\arraybackslash}p{2cm}|
                    >{\centering\arraybackslash}p{2cm}|
                    }
        \hline 
         Ref &  Multiple Datasets & Diversity of Attack Types & FL Implementation &  FL Parameter Optimization & XAI Integration & Global Explanations & Ethical Discussion  \\ \hline 
         \cite{al2023privacy}         & \checkmark & X & X & X & X & X & X \\ \hline 
         \cite{otoum2021federated}    & X & X & \checkmark & X & X & X & X \\ \hline 
         \cite{schneble2019attack}    & X & \checkmark & \checkmark & \checkmark & X & X & \checkmark \\ \hline 
         \cite{singh2022dew}          & \checkmark & \checkmark & \checkmark & X & X & X & X \\ \hline 
         \cite{tanveer2022xsru}       & X & \checkmark  & X & X & \checkmark & X & X \\ \hline 
         \cite{sun2024optimized}      & X & \checkmark & X & X & X & X & \checkmark \\ \hline 
         \cite{kilincer2023automated} & \checkmark & \checkmark & X & X  & X & X & X \\ \hline 
         \cite{gu2023intrusion}       & X & X & X & X  & X & X & X \\ \hline 
         \cite{ravi2023deep}          & \checkmark & \checkmark & X & X  & X & X & X \\ \hline
         \cite{chaganti2022particle}  & X & X & X & X  & X & X & X \\ \hline
         \cite{alalhareth2023improved}& X & X & X & X  & X & X & X \\ \hline
         \cite{rm2020effective}       & \checkmark & \checkmark & X & X & X & X & X \\ \hline
         \cite{kumar2021ensemble}     & X & \checkmark & X & X & X & X & X \\ \hline
         \cite{gupta2022cybersecurity}& \checkmark & \checkmark & X & X & X & X & X \\ \hline
         \cite{hady2020intrusion}     & X & X & X & X & X & X & X \\ \hline
         Proposed Framework&  \checkmark &  \checkmark & \checkmark & \checkmark & \checkmark & \checkmark & \checkmark \\ \hline
    \end{tabular}
    \caption{Comparison of the Proposed Approach with Previous Works } \label{tbl2} 
    \label{tab:Comparison_prev_works}
\end{table}

\section{DISCUSSION} \label{DISCUSSION}

The proposed framework for intrusion detection in the IoMT  system is based on DL for anomaly detection, FL for model training and privacy protection, and XAI  for enhanced interpretability and explainability. This distinctive approach, in contrast to existing literature, sets our work apart.
The introduction of an IDS  based on DL underscores its real-time efficiency in swiftly identifying network or host-based attacks that could compromise the integrity, confidentiality, and availability of the IoMT system.
By leveraging DL as a ML method, our framework enables the automatic selection of pertinent features, offering an end-to-end solution that operates seamlessly without relying on third-party interventions. This streamlined approach enhances the robustness and autonomy of the intrusion detection process within the IoMT system.

The proposed solution places a paramount emphasis on safeguarding patient privacy by employing FL, which involves sharing model weights instead of actual patient data. By optimizing FL parameters, the communication rounds are significantly reduced, thereby minimizing bandwidth consumption, preventing network congestion, and ensuring the scalability of the system.
The distributed nature of FL proves instrumental in positioning the IDS  in close proximity to potential attack sources. This proximity enhances the system's agility, allowing for rapid and effective detection and response to security threats. This decentralized approach strengthens the overall security posture while contributing to the rapid identification and mitigation of potential risks to patient data within the IoMT ecosystem.

The proposed system provides an explanation and interpretation of the ML model for anomaly detection, thereby enhancing trust in the capability of the proposed DL-based IDS for anomaly detection. Concurrently, it assists regulators seeking to verify compliance with international standards. For users of the proposed framework, such as patients, trust can be reinforced by presenting a performance history achieved by the system. This historical record serves to strengthen trust in the reliable execution of the system.
Moreover, the proposed IDS functions as a valuable decision aid, empowering CISO to intervene promptly in case of anomaly detection. Whether triggered by communication issues, medical emergencies, or security attacks, this intervention capability ensures a proactive response to safeguard the integrity and security of the IoMT system.

Demonstrating high predictive performance across multiple benchmark datasets, encompassing both network traffic and medical data, provides strong evidence of the effectiveness and robustness of the proposed framework. Beyond conventional evaluation metrics, the comparative analysis between centralized and FL further demonstrates that both approaches rely on highly consistent feature attribution patterns for anomaly detection. The SHAP analysis reveals that the same domain-relevant network and medical features remain the primary contributors to the classification process, exhibiting comparable attribution patterns under both training paradigms. These findings suggest that decentralized optimization preserves the feature attribution patterns of the centralized model without introducing artificial feature dependencies. More importantly, this comparison highlights the practical contribution of XAI within the proposed framework. Without SHAP, the comparison between centralized and FL would be limited to predictive performance metrics, making it impossible to determine whether both models reach their predictions based on the same evidence. By showing that the federated model consistently relies on the same clinically and network-relevant features as the centralized model, with comparable attribution patterns, XAI provides complementary evidence that privacy-preserving learning maintains the interpretability and reliability of the intrusion detection process. Consequently, the proposed FL-based IDS not only achieves detection performance comparable to centralized learning while preserving data privacy, reducing communication overhead through optimized federated training, and mitigating single points of failure, but also provides an interpretable and trustworthy decision-support framework suitable for security-critical IoMT environments.

The proposed framework makes a significant ethical contribution by addressing key concerns related to data privacy, transparency, fairness, and accountability in the context of IoMT systems. By leveraging FL, the framework ensures that sensitive patient data remains on local devices, thereby preserving privacy and complying with stringent regulations such as HIPAA and GDPR. This approach minimizes the risk of data breaches and unauthorized access, fostering trust among patients and healthcare providers. Furthermore, the integration of XAI methods, such as SHAP, enhances the transparency of the decision-making process, allowing stakeholders to understand and verify the model's predictions. This transparency is crucial for ensuring accountability, as it enables the identification and mitigation of potential biases or errors in the system. Additionally, the framework promotes fairness by using diverse datasets and mitigating biases, ensuring that the benefits of the technology are accessible to all patients without discrimination. These ethical considerations are essential for building trust and ensuring the responsible deployment of AI in healthcare, ultimately contributing to the well-being and safety of patients.

Yet, a significant challenge emerges due to the scarcity of datasets explicitly tailored for IoMT  systems, featuring diverse attacks and balanced class instances. This scarcity impedes the validation of security solutions and the comparative analysis of distinct contributions to IoMT system security. Additionally, deploying these solutions in real-world environments introduces unforeseen challenges, such as dynamic network conditions, device heterogeneity, and unpredictable user behavior. Addressing these complexities will be essential to developing robust, scalable, and practical security frameworks for IoMT systems.

\section{CONCLUSION} \label{CONCLUSION}

In conclusion, this study introduces a framework for an IDS based on an ANN enhanced with FL and XAI methods. The synergistic integration of these components enhances robust attack detection in the context of the IoMT, emphasizing data privacy and ensuring model explainability and interpretability. The IDS architecture capitalizes on FL, fostering collaborative model training while upholding the confidentiality of sensitive data, thereby addressing privacy concerns prevalent in healthcare. Furthermore, the FL process is designed to accommodate the operational characteristics of IoMT environments, including dynamic client participation, heterogeneous device resources, and communication constraints. Additionally, the incorporation of XAI bolsters transparency, ensuring compliance with regulatory requirements and healthcare legislation. This, in turn, cultivates greater trust in the decision-making process of the system among stakeholders.
A comprehensive evaluation across diverse datasets containing both network and medical data underscores the applicability and resilience of the proposed solution. The experimental analysis further demonstrates how key FL parameters, including the number of participating clients, the client participation fraction, and the number of local training epochs, influence communication overhead, convergence behavior, and intrusion detection performance under representative IoMT operating conditions. Particularly noteworthy are the results showcasing the efficacy of the proposed FL framework, achieving an accuracy surpassing 98\%, comparable to traditional centralized approaches. Furthermore, the provision of explanations and result interpretations using XAI adds an extra layer of assurance, reinforcing trust for ML model designers, regulators, and users of IoMT systems within the proposed framework.
Beyond its technical achievements, the proposed framework makes a significant ethical contribution by ensuring the protection of sensitive patient data through FL, promoting transparency and accountability via XAI, and fostering fairness by mitigating biases in the model. By aligning with international standards such as HIPAA, GDPR, WHO and ISO/IEC 27701, the framework not only enhances the security of IoMT systems but also ensures that the deployment of AI in healthcare is ethical, responsible, and equitable. These ethical considerations are integral to building trust among patients, healthcare providers, and regulators, ultimately contributing to the safe and effective use of AI in healthcare.

Future work will aim to fortify the solution against emerging threats such as poisoning attacks, adversarial ML, and quantum attacks, while optimizing the FL model via advanced client selection and dataset partitioning strategies. Additionally, privacy-preserving approaches, including differential privacy and secure multi-party computation, will be further explored to enhance data confidentiality and security. The solution will undergo extensive real-world testing to identify and address practical challenges, ensuring its robustness and scalability in diverse environments.

\section{CRediT authorship contribution statement}
\textbf{Ayoub Si-ahmed:} Methodology, Investigation, Writing - Original Draft
\textbf{Mohammed Ali Al-Garadi:} Methodology, Writing - Review \& Editing, Supervision, Project administration.
\textbf{Narhimene Boustia:} Resources, Writing - Review \& Editing, Supervision, Project administration.

\section{Declaration of Competing Interest}
The authors declare that they have no known competing financial interests or personal relationships that could have influenced the work reported in this paper.

\bibliographystyle{unsrt}
\bibliography{conf.bib}

\end{document}